\documentclass[amsmath,amssymb,bm]{pinchcr}   
 \usepackage{makeidx} 
\usepackage{amsmath}
\usepackage{bm}
\usepackage[dvips]{graphicx}
\usepackage{amssymb}

\newcommand{\QS}{Q_{\mathcal S}}
\newcommand{\PS}{P_{\mathcal S}}

\newcommand{\ep}{\varepsilon}
\newcommand{\M}{_{\rm M}}

\newcommand{\ad}{a^{\dagger}}
\newcommand{\lal}{{\langle\langle}}
\newcommand{\rar}{{\rangle\rangle}}

\addtocounter{chapter}{6}

\begin{document}


\maintext

\chapter[Periodically modulated quantum nonlinear oscillators]{Periodically modulated quantum nonlinear oscillators\\
{\normalsize M. I. Dykman}}

\section{Introduction}
\label{sec:dykman_intro}

\begin{figure}[htbp]
\begin{center}
\includegraphics[width=4.cm]{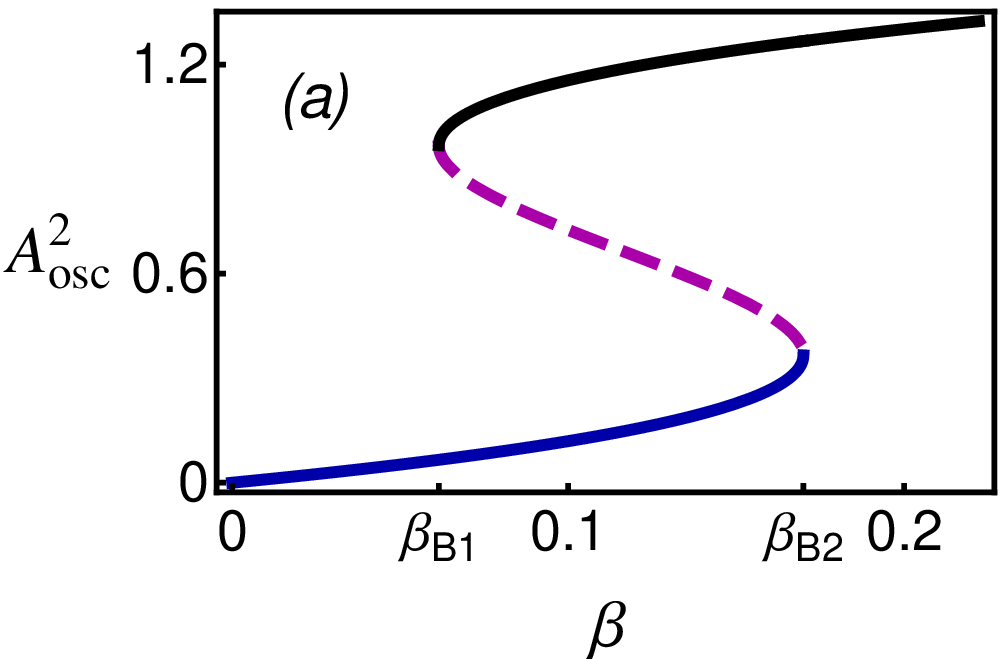}\hspace*{0.5in}
\includegraphics[width=3.9cm]{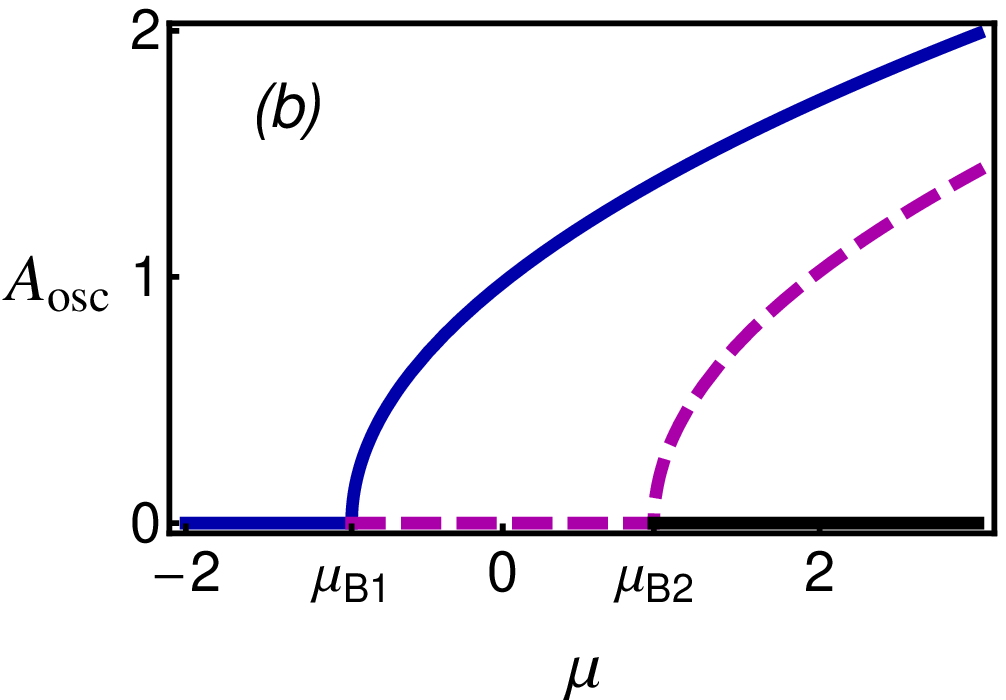}
\end{center}
\caption{\index{bistability}Bistability of forced vibrations for additive (a) and parametric (b) modulation at frequency $\omega_F$ close to the oscillator eigenfrequency $\omega_0$ and $2\omega_0$, respectively; $A_{\rm osc}$ is the scaled vibration amplitude. In (a), $\beta$ is the scaled squared modulation amplitude, and in (b) $\mu$ is the scaled frequency detuning $\omega_F-2\omega_0$. Parameters $\beta,\mu$ and the scaling factor of the vibration amplitude $C$ are defined in Table~\ref{table:parameters}. Solid and dashed lines show the stable and unstable stationary vibrational states; in (b) there are two states with the same nonzero $A_{\rm osc}$ and the phases that differ by $\pi$. The bifurcation parameter values $\beta_{B1,2}$, $\mu_{B1,2}$ indicate where the number of stable vibrational states changes. The scaled decay rate $\kappa$ in (a) and (b) is 0.25 and 0.3, respectively.}
\label{fig:bistability}
\end{figure}

Vibrational systems have been attracting much attention in physics. Such systems are always nonlinear, at least to some extent. For weak damping, even small nonlinearity can become important. For example, classically, the nonlinearity-induced dependence of the vibration frequency on amplitude can lead to bistability of forced resonant vibrations \shortcite{LL_Mechanics2004}, see Fig.~\ref{fig:bistability}. Quantum mechanically, the nonlinearity
makes the frequencies of transitions between adjacent energy levels different and thus enables spectroscopic identification and selective excitation of these transitions.
The interest in quantum effects in oscillators significantly increased recently in the context of nonlinear vibrations in Josephson junction based systems and applications of these systems in quantum information \shortcite{Wallraff2004,Siddiqi2005,Lupascu2006,Steffen2006,Metcalfe2007,Schreier2008,Watanabe2009,Mallet2009,Vijay2009,Wilson2010,Bishop2010,Reed2010}.  The long-sought \shortcite{Blencowe2004a,Schwab2005a} quantum regime has been reached also in nanomechanical resonators \shortcite{O'Connell2010,Riviere2011}. This development makes it possible to study quantum effects in individual vibrational systems rather than ensembles.

Besides being interesting on their own, nonlinear oscillators allow addressing some fairly general physics problems. One of them is classical and quantum fluctuations far from thermal equilibrium and whether they have features that have no analog in systems close to equilibrium. A resonantly modulated nonlinear oscillator could be the first well-characterized physical system  with no detailed balance, which was used to study an important class of fluctuation phenomena, the fluctuation-induced switching between coexisting stable states, both in the classical and quantum regimes, and to reveal some of such features\footnote{For the theory and experiment on switching of a resonantly modulated oscillator with no detailed balance see, in particular, \shortciteNP{Dykman1979a,Dmitriev1986a,Vogel1990,Dykman1998,Lapidus1999,Siddiqi2005,Kim2005,Aldridge2005,Stambaugh2006,Almog2007} for the classical and \shortciteNP{Dykman1988a,Vogel1988,Kinsler1991,Marthaler2006,Katz2007,Serban2007,Vijay2009,Mallet2009,Peano2010,Wilson2010}, for the quantum regime.}.

An important source of quantum fluctuations in an oscillator is coupling to a thermal bath. The coupling leads to oscillator relaxation via emission of excitations in the bath (photons, phonons, etc) accompanied by transitions between the oscillator energy levels. If the coupling is weak, the transition rates are small compared to the energy transferred in a transition, in frequency units. In the classical limit, the energy levels are not resolved and the transitions lead to friction.

At the quantum level, one should take into account that the transitions happen at random. The randomness gives rise to a peculiar quantum noise. For a resonantly modulated oscillator, it leads to diffusion over the quantum states, which are time-dependent because of the modulation. The result of this diffusion is quantum heating of the oscillator. It is seen, in particular, in a nonzero {\em width} of the distribution over the oscillator states even where the temperature of the thermal reservoir is zero. Quantum heating is qualitatively different from the familiar Joule heating, which characterizes the power absorbed from the modulating field. In contrast to the Joule heating, the resulting distribution over the states turns out to be independent of the oscillator relaxation rate, for weak damping.

A consequence of quantum heating is quantum activation \shortcite{Dykman1988a,Marthaler2006,Dykman2007,Katz2007,Peano2010a}. This is a mechanism of switching between coexisting stable states of forced vibrations shown in Fig.~\ref{fig:bistability}. Similar to quantum heating, the switching is due to the quantum noise that accompanies relaxation. It occurs via transitions over an effective barrier that separates the vibrational states. The mechanism differs from quantum tunneling, and moreover, leads to an exponentially larger switching rate even for low temperatures. It differs also from thermal activation, which becomes important in the classical regime of high temperatures.

Quantum activation has no analog in systems close to thermal equilibrium. The switching rate has a characteristic dependence on the parameters and displays scaling behavior with characteristic exponents. This has made it possible to identify quantum activation in the experiment \shortcite{Vijay2009}. The physics of quantum heating and quantum activation is explained in Secs.~\ref{sec:resonant_modulation} and \ref{sec:quantum_activation}, respectively.

Quantum heating is manifested also in the power spectra of resonantly modulated oscillators and the spectra of their response to a weak additional field. Spectroscopy has been recognized as a means of studying the dynamics of modulated oscillators and, more recently, of using oscillators for quantum measurements \shortcite{Dykman1979a,Drummond1980c,Collett1985,Dykman1994b,Stambaugh2006a,Chan2006,Nation2008,Wilson2010,Vierheilig2010,Boissonneault2010,Laflamme2011}.   Because of the interplay of many interstate transitions at close frequencies, the spectra of modulated quantum oscillators have a characteristic shape. They can have a fine structure, in which case they directly provide the quantum-heating induced distribution over the oscillator states   \shortcite{Dykman2011}. The power spectra are also important  for understanding the dynamics of resonantly driven oscillators coupled to two-level systems \shortcite{Picot2008,Serban2010}; interesting spectral manifestations of quantum heating in such coupled systems have been recently found\footnote{F.~R.~Ong {\textit et al.}, in preparation (experiment) and M. Boissonneault {\textit et al.}, in preparation (theory); we are grateful to P. Bertet for informing us about this work}. The spectra of modulated oscillators are discussed in Sec.~\ref{sec:q_heat_spectra}.

Nonresonant modulation can also have pronounced effect on the oscillator dynamics. Recently such modulation attracted much attention in optomechanics, where intracavity modes are coupled to mechanical vibrations, for example, to the vibrations of a mirror in the cavity \shortcite{Kippenberg2008}.
Modulation can lead to cooling and  heating of an oscillator, or excite self-sustained vibrations. In contrast to the quantum heating discussed above, where quantum fluctuations broaden the distribution over the oscillator states in a strong resonant field, here the issue is the change of the distribution over the Fock states of the oscillator. An interesting feature of heating and cooling is that, in the important case where the effective friction is linear (as for standard viscous friction), the distribution over the Fock states remains of the Boltzmann form, but the temperature differs from the bath temperature \shortcite{Dykman1978,Clerk2004a,Wilson-Rae2007,Marquardt2007}. In Sec.~\ref{sec:cooling} we extend the previous analysis \shortcite{Dykman1978} to take into consideration both the nonlinearity of the coupling of the oscillator to other degrees of freedom and the nonlinearity of the interaction with the modulating field.

\section{Resonant modulation: Quantum heating}
\label{sec:resonant_modulation}

For weakly damped oscillators, vibration nonlinearity becomes important once the change of the vibration frequency due to the nonlinearity $\Delta\omega$ becomes comparable to the oscillator decay rate  $\Gamma$, which characterizes frequency uncertainty. This happens where $\Delta\omega$ is still small compared to the oscillator eigenfrequency $\omega_0$ and the nonlinear part of the vibration energy is small compared to the harmonic part. Respectively, the vibrations remain almost sinusoidal, which significantly simplifies the analysis. At the same time, for weak damping an already moderately strong resonant modulation can drive an oscillator into the amplitude range where $\Delta\omega\gtrsim \Gamma$. This makes underdamped oscillators advantageous for studying quantum phenomena far from equilibrium.

\subsection{Oscillator Hamiltonian in the rotating frame}
\label{subsec:Hamiltonian}

The most frequently used types of \index{modulation}\index{modulation!resonant}resonant modulation of an oscillator are modulation by a resonant additive force $A\cos\omega_Ft$ with frequency $\omega_F$ close to $\omega_0$ and parametric modulation by force $F\cos\omega_Ft$ with $\omega_F$ close $2\omega_0$. For moderately strong modulation it is often sufficient to take into account only the leading-order oscillator nonlinearity which leads to the amplitude dependence of the vibration frequency. It is sometimes called Kerr nonlinearity, and the corresponding model of the oscillator is called the Duffing model.\index{Duffing oscillator} The Hamiltonian of the Duffing oscillator is
\begin{eqnarray}
\label{eq:H_0(t)}
H_0=\frac{1}{2}p^2+\frac{1}{2}\omega_0^2q^2 +\frac{1}{4}\gamma q^4 + H_F(t),
\end{eqnarray}
where $q$ and $p$ are the oscillator coordinate and momentum, the mass is set equal to one, and
$\gamma$ is the anharmonicity parameter.

The modulation term $H_F$ in eq.~(\ref{eq:H_0(t)}) for additive ($H_F= H_{\rm add}$) \index{modulation!additive}and \index{modulation!parametric}parametric ($H_F= H_{\rm par}$) modulation has the form
\begin{equation}
\label{eq:modulation_hamiltonian}
H_{\rm add}=-qA\cos \omega_Ft,\qquad H_{\rm par}=\frac{1}{2}q^2F\cos \omega_Ft.
\end{equation}
The conditions that the modulation is resonant and not too strong are
\begin{equation}
\label{eq:delta_omega}
|\delta\omega|\ll \omega_0, \qquad \delta\omega=\omega_{\rm M}-\omega_0; \qquad |\gamma|\langle q^2
\rangle\ll\omega_0^2.
\end{equation}
Here, $\omega_{\rm M}$ is equal to $\omega_F$ and $\omega_F/2$ for additive and parametric modulation, respectively; this is the frequency close to the oscillator eigenfrequency for these types of resonant modulation.  For concreteness, we assume $\gamma, F>0$; for additive driving, the oscillator can be bistable for $\gamma\delta\omega > 0$; we assume $\delta\omega>0$ for such driving.

It is convenient to change to the rotating frame using the standard canonical transformation $U(t)=\exp\left(-ia^{\dag}a\,\omega_{\rm M}t\right)$, where $a^{\dag}$ and $a$ are the raising and lowering operators of the oscillator. We introduce slowly varying in time dimensionless coordinate $Q$ and momentum $P$, using as a scaling factor the characteristic amplitude of forced vibrations $C$, see Table~\ref{table:parameters},
\begin{eqnarray}
\label{eq:canonical_transform}
U^{\dag}(t)q U(t) &=& C_{}(Q\cos\varphi_{}+P\sin\varphi_{}), \nonumber\\
U^{\dag}(t)p U(t) &=& -C_{}\omega_{\rm M}(Q\sin\varphi_{} - P\cos\varphi_{}).
\end{eqnarray}
For additive and parametric modulation $\varphi_{} \equiv \varphi_{\rm add}=\omega_Ft$ and $\varphi_{} \equiv\varphi_{\rm par}=(\omega_Ft+\pi)/2$, respectively. The commutation relation between $P$ and $Q$ has the form
\begin{equation}
\label{eq:lambda}
[P,Q]=-i\lambda,\qquad \lambda=\hbar/(\omega\M C_{}^2) .
\end{equation}
Parameter $\lambda\propto \hbar$ plays the role of the Planck constant in the quantum dynamics in the rotating frame. It is determined by the oscillator nonlinearity, $\lambda\propto \gamma$, see Table~\ref{table:parameters}.
For characteristic $|Q|,|P|\lesssim 1$, where $\langle q^2\rangle \lesssim C^2$, the last inequality in eqn~(\ref{eq:delta_omega}) coincides with the first inequality in this equation for additive modulation, whereas for parametric modulation is gives condition $F\ll \omega_0^2$.
\noindent
\begin{table}[htbp]
\tableparts
{
\caption{Parameters of a resonantly modulated oscillator}
\label{table:parameters}
}
{\begin{tabular}{lll}
\hline
 & & \\[-8pt]
& Additive driving & Parametric driving\\ [3pt]
\hline
 & &\\[-5pt]
Amplitude scale & $C=\left[8\omega_F(\omega_F-\omega_0)/3\gamma\right]^{1/2}
$ & $ C=|2F/3\gamma|^{1/2}$\\[5pt]
Scaled Planck constant &
$\lambda = 3\hbar\gamma/8\omega_F^2(\omega_F-\omega_0)$ &
$\lambda = 3\hbar\omega_F^{-1}|\gamma/F|$\\[5pt]
Control parameter & $\beta=3\gamma A^2/32\omega_F^3(\omega_F-\omega_0)^3 $ & $\mu= \omega_F(\omega_F-2\omega_0)/|F|$\\[5pt]
Scaled decay rate$^*$ & $\kappa=\Omega^{-1}=\Gamma/|\omega_F-\omega_0|$ &
$\kappa=\zeta^{-1}=2\Gamma\omega_F/|F|$\\[5pt]
\hline
\end{tabular}
}
{\footnotesize $^*$ Notations $\Omega$ and $\zeta$ were used in some of our previous papers, cf. \shortciteN{Dykman2007}.}
\end{table}

In the range (\ref{eq:delta_omega}) the oscillator dynamics can be analyzed in the \index{rotating wave approximation}rotating wave approximation (RWA). The Hamiltonian in the rotating frame\index{rotating frame} is
\begin{equation}
\label{eq:H_rotating_frame}
\tilde H_0=U^{\dag}H_0 U-i\hbar U^{\dag}\dot{U}\approx
(3E_{\rm sl}/8)\hat{g}_{}, \qquad E_{\rm sl}=\gamma C_{}^4,
\end{equation}
where $E_{\rm sl}\sim \gamma\langle q^4\rangle$ is the characteristic energy of motion in the rotating frame. This motion is slow on the time scale $\omega_F^{-1}$. Operator $\hat g_{}=g(Q,P)$ in eqn~(\ref{eq:H_rotating_frame}) is independent of time. For additive and parametric modulation, respectively, we have
\begin{eqnarray}
\label{eq:g_additive}
g_{\rm add}(Q,P) =
\frac{1}{4}\left(P^2+Q^2 -1\right)^2 - \beta^{1/2}Q,
\end{eqnarray}
and
\begin{equation}
\label{eq:g_parametric}
g_{\rm par}(Q,P) =
\frac{1}{4}\left(P^2+Q^2\right)^2
+\frac{1}{2}(1-\mu)P^2
-\frac{1}{2}(1+\mu)Q^2.
\end{equation}
\noindent
\begin{figure}[h]
\begin{center}
\includegraphics[width=3.5cm]{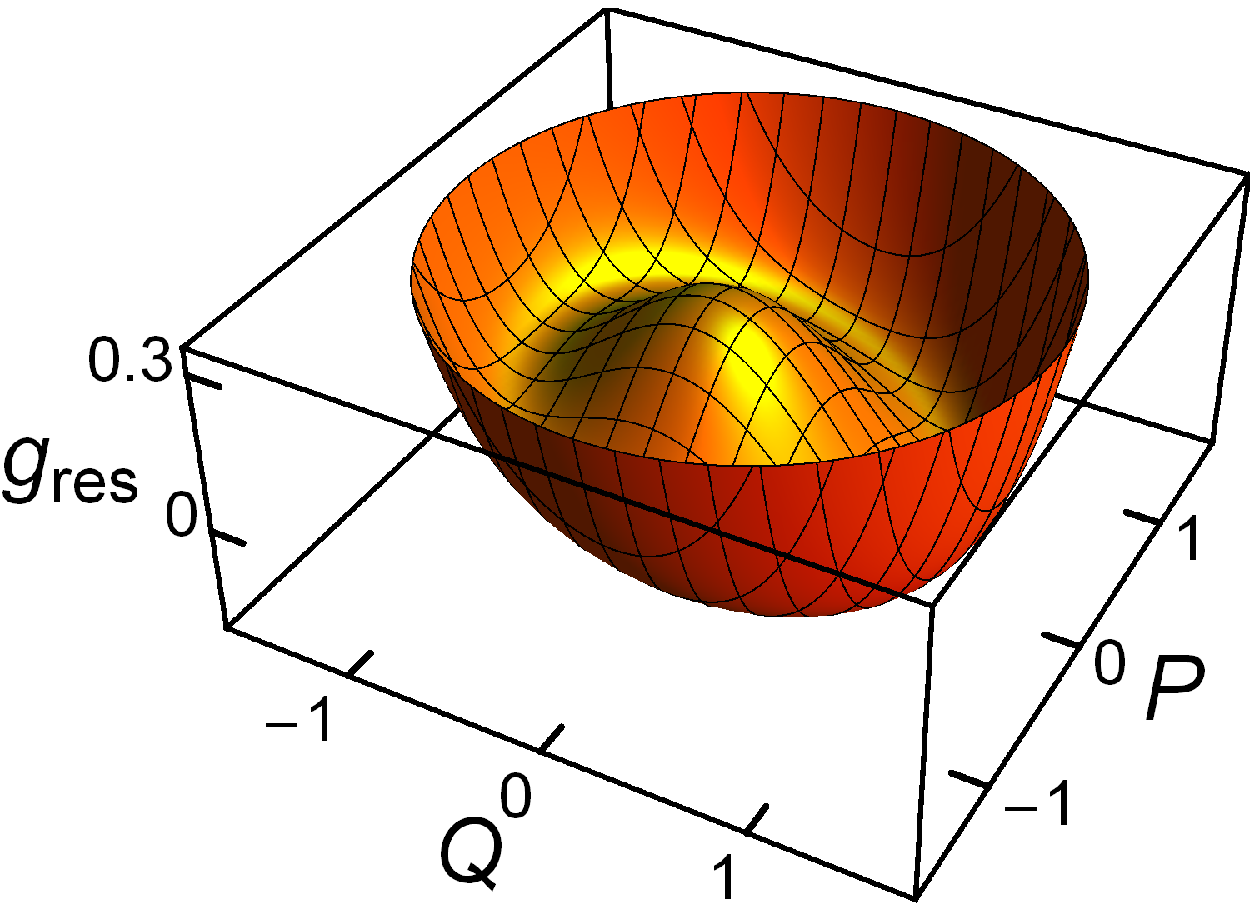}\hspace*{0.5in}
\includegraphics[width=3.5cm]{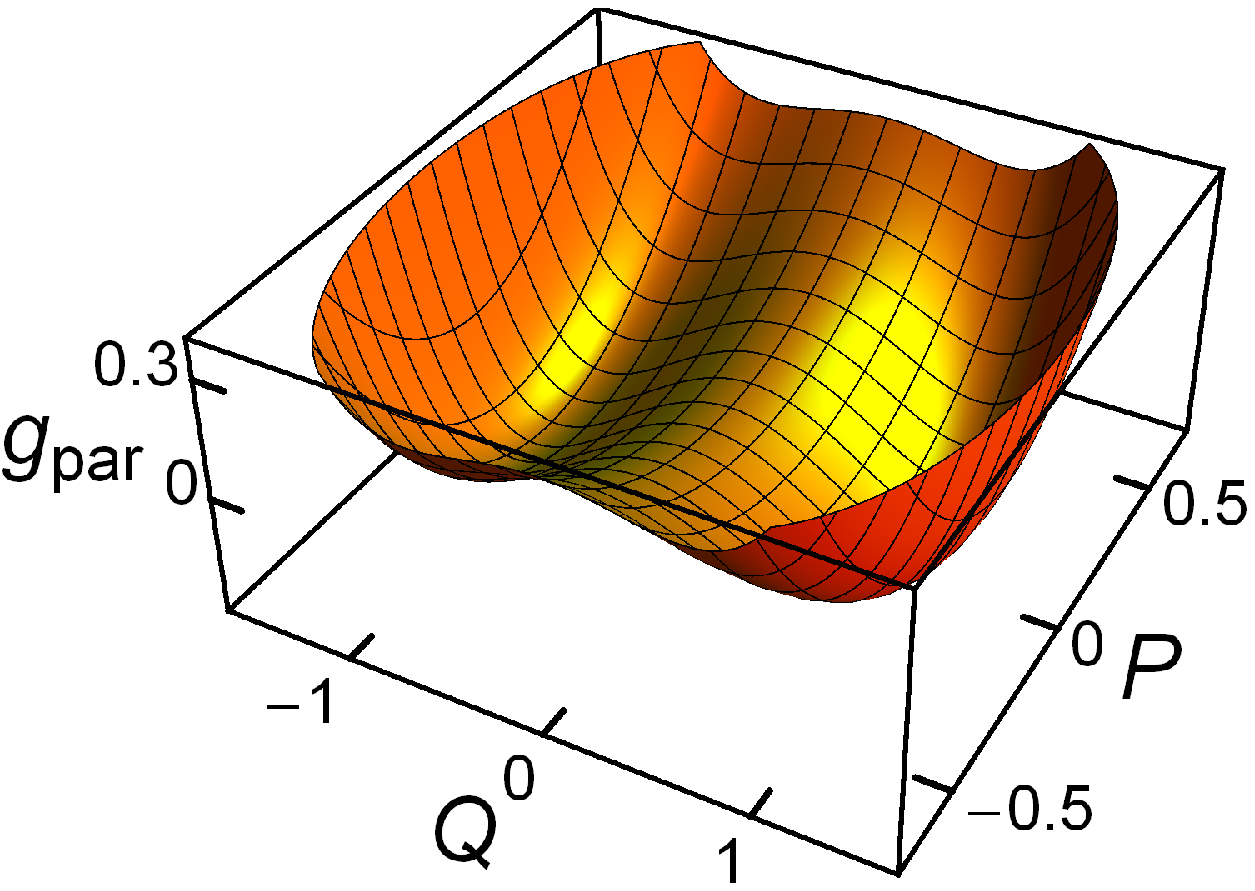}
\end{center}
\caption{(Color) The dimensionless Hamiltonian functions of the oscillator for additive (left panel) \index{modulation!resonant}\index{modulation!parametric}and parametric (right panel) modulation. The plots refer, respectively, to $\beta = 0.01$  and to $\mu=-0.1$. In the presence of weak dissipation, the minimum and the local maximum of $g_{\rm add}$ and the minima of $g_{\rm par}$ become classically stable states of forced vibrations in the lab frame.}
\label{fig:quasienergy_surfaces}
\end{figure}
In Fig.~\ref{fig:quasienergy_surfaces} we show $g_{\rm add}(Q,P)$ and $g_{\rm par}(Q,P)$ as functions of classical coordinate and momentum.  Each of these functions depends on one dimensionless parameter, $\beta$ and $\mu$, respectively, which characterizes the ratio of the modulation strength to the frequency detuning. These parameters are given in Table~\ref{table:parameters}.
For $0<\beta<4/27$ function $g_{\rm add}$ has the form of a tilted Mexican hat, with a local maximum and with a minimum at the lowest point of the rim. In the presence of weak dissipation these extrema correspond to classically stable states of forced vibrations with small and large amplitude, respectively, see Fig.~\ref{fig:bistability}(a).

For $-1 < \mu < 1$, function $g_{\rm par}$ has two minima, which in the presence of weak dissipation correspond to stable vibrational states,\index{stable state} see Fig.~\ref{fig:bistability}(b). Function $g_{\rm par}$ has symmetry $g_{\rm par}(Q,P)= g_{\rm par}(-Q, -P)$. This is a consequence of the time-translation symmetry $H_0(t)=H_0(t+2\pi/\omega_F)$, as seen from eqn~(\ref{eq:modulation_hamiltonian}) and (\ref{eq:canonical_transform}). Respectively, the vibration amplitudes in the stable states are the same, but the vibration phases differ by $\pi$, characteristic of parametric resonance.

\subsection{Quasienergy spectrum}
\label{subsec:quasienergies}

Operator $\hat g$ plays the role of dimensionless Hamiltonian of the modulated oscillator in the rotating frame. In the RWA, the Schr\"odinger equation in dimensionless slow time $\tau$ reads
\begin{equation}
\label{eq:Schrodinger_eq}
i\lambda\dot\psi \equiv i\lambda \partial_{\tau}\psi= \hat g \psi, \qquad \tau= t\lambda\gamma C^4/\hbar\equiv (\lambda E_{\rm sl}/\hbar)t;
\end{equation}
$\tau = t\,\delta\omega$ and $\tau = tF/2\omega_F$ for additive and parametric modulation, respectively.

Operator $\hat g$ has a discrete spectrum, $\hat g|n\rangle=g_n|n\rangle$. The eigenvalues $g_n$ have simple physical meaning. A periodically modulated oscillator does not have stationary states with conserved energy in the lab frame. It is rather described by the Floquet, or \index{quasienergy}quasienergy states $\Psi_{\ep}(t)=U(t)\psi_{\ep}(\tau)$, which is a consequence of the periodicity of the Hamiltonian $H_0(t)=H_0(t+2\pi/\omega_F)$. One can seek a solution of the full Schr\"odinger equation $i\hbar\partial_t\Psi=H_0(t)\Psi$ in the form $\Psi_{\ep}(t+t\M)=\exp(-i\ep t\M/\hbar)\Psi_{\ep}(t)$, where $t\M=2\pi/\omega\M$ ($t\M=2\pi/\omega_F$ and $t\M=4\pi/\omega_F$ for additive and parametric modulation, respectively). This expression defines quasienergy $\ep$. We note that, for parametric modulation, we use the doubled modulation period when defining $\ep$; this is convenient for the description of period-two states of the oscillator.

\begin{figure}[h]
\begin{center}
\includegraphics[width=7cm]{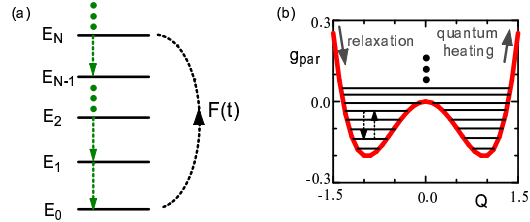}
\end{center}
\caption{Panel (a): oscillator relaxation and excitation. Relaxation is due to transitions between the Fock states with energies $E_N\approx \hbar\omega_0(N+1/2)$ accompanied by emission (or absorption, for nonzero temperature) of excitations in the bath, e.g., photons. The stationary vibrational state is formed on balance between relaxation and excitation by periodic modulation $F(t)$. Panel (b): the effective RWA Hamiltonian of the parametrically modulated oscillator  $g_{\rm par}(Q,P=0)$ for $\mu=-0.1$, with sketched quasienergy levels. The dashed arrows indicate transitions between quasienergy states, which are due to emission of excitations in the bath. The solid arrows indicate the change of $g$ in relaxation and quantum heating.}
\label{fig:relaxation_sketch}
\end{figure}

From eqns~(\ref{eq:modulation_hamiltonian}) and (\ref{eq:Schrodinger_eq}), in the RWA  the oscillator quasienergies are simply related to the eigenvalues of $\hat g$, $\ep_n= (3E_{\rm sl}/8)g_n$, i.e., $g_n$ is a scaled quasienergy.\index{quasienergy!spectrum} Here we are using an extended $\ep$-axis rather than limiting $\ep$ to the analog of the first Brillouin zone $0\leq \ep<\hbar\omega_{\rm M}$.
The scaled quasienergy spectrum for parametric modulation in the neglect of tunneling is sketched in the right panel of Fig.~\ref{fig:relaxation_sketch}. In the region of bistability,\index{bistability} $-1 < \mu < 1$, the states with $g_n<0$ are degenerate. Function $g_{\rm par}(Q,P=0)$ has a form of a symmetric double-well potential, and the states with $g_n<0$ remind intrawell states of a particle in such potential. Semiclassically, in the rotating frame the oscillator moves along orbits, which are cross-sections of the surface $g_{\rm par}$ in Fig.~\ref{fig:quasienergy_surfaces} by planes $g_{\rm par}=g_n$.

The structure of quasienergy states of an additively driven oscillator can be understood in a similar way by thinking of the cross-sections of the surface $g_{\rm add}(Q,P)$ in Fig.~\ref{fig:quasienergy_surfaces} by planes $g_{\rm add}=g_n$. The eigenstates localized near the local maximum of $g_{\rm add}(Q,P)$ correspond to semiclassical orbits on the surface of the ``inner dome" of  $g_{\rm add}(Q,P)$; these states become stronger localized as $g_n$ {\em increases} toward the local maximum of $g_{\rm add}(Q,P)$. This is in contrast with the conventional picture of a particle in a potential well, where the localization becomes stronger  with decreasing energy.
\noindent

Many interesting and unusual features of the dynamics described by eqn~(\ref{eq:Schrodinger_eq}) follow from the fact that functions $g_{\rm add},g_{\rm par}$ do not have the form of a sum of the kinetic and potential energies. These features are seen, in particular, in tunneling,\index{tunneling} which is significantly modified compared to the conventional picture, because for a given $g(Q,P)$ the momentum $P$ as function of coordinate $Q$ has 4 rather than 2 branches \shortcite{Dmitriev1986,Serban2007}. One of the consequences is that, for example, for a parametric oscillator, decay of the wave function in the classically inaccessible region of $Q$ can be accompanied by oscillations, leading to under-barrier interference  \shortcite{Marthaler2007a}.

As seen from eqn~(\ref{eq:Schrodinger_eq}), the quasienergy level spacing is $\propto \lambda E_{\rm sl}$. From eqn~(\ref{eq:delta_omega}), it is small compared to the distance between the oscillator energy levels in the absence of modulation,
$|\ep_n-\ep_{n+1}|\sim \lambda E_{\rm sl} \ll \hbar\omega_0$.


\subsection{Qualitative picture of quantum heating and the master equation}
\label{subsec:qualitative_heating}

\index{quantum heating}Quantum heating is most easy to understand in the case where the oscillator decay rate $\Gamma$\index{decay rate} is small not only compared to $\omega_0$, but also to the distance between quasienergy levels in frequency units, $\Gamma\ll \lambda E_{\rm sl}/\hbar$. In this Section we consider relaxation due to coupling to a bosonic thermal bath with the coupling Hamiltonian $H_i$ linear in the oscillator coordinate and momentum and thus in the oscillator ladder operators $a, a^{\dag}$ \shortcite{Schwinger1961},
\begin{equation}
\label{eq:linear_relaxation}
H_i= ah_{\rm b}+{\rm H.c.},\qquad \Gamma\equiv\Gamma(\omega_0)=\hbar^{-2}{\rm Re}~\int\nolimits_0^{\infty}dt\langle [h_{\rm b}^{\dag}(t),h_{\rm b}(0)]\rangle_{\rm b}e^{i\omega_0t},
\end{equation}
where $h_{\rm b}$ depends on the bath variables only and $\langle\ldots\rangle_{\rm b}$ denotes thermal averaging over the bath states; in what follows we assume $\langle h_{\rm b}\rangle_{\rm b} = 0$.

For the interaction (\ref{eq:linear_relaxation}), in the absence of modulation relaxation is due to transitions between adjacent oscillator Fock states $|N\rangle$.  For zero bath temperature, these transitions occur only downward in energy, with emission of excitations in the bath, see the left panel of Fig.~\ref{fig:relaxation_sketch}.  For a smooth density of states of the bath, resonant modulation does not change the decay rate, $\Gamma(\omega_{\rm M})\approx \Gamma(\omega_0)$. However, it excites the oscillator, as sketched in Fig.~\ref{fig:relaxation_sketch}. In the stationary vibrational state the energy provided by the modulation is balanced by relaxation.

As outlined in Sec.~\ref{sec:dykman_intro}, the randomness of the transitions, i.e., the quantum noise that accompanies relaxation, leads to quantum heating. The heating can be understood from Fig.~\ref{fig:relaxation_sketch} by noticing that the quasienergy states $|n\rangle$ ($\hat g|n\rangle = g_n|n\rangle$) sketched in the right panel are linear combinations of the Fock states $|N\rangle$ in the left panel, $|n\rangle = \sum\nolimits_N a_{nN}|N\rangle$. Therefore transitions between the Fock states downward in oscillator energy correspond to transitions both downward and upward in quasienergy, with different rates. If the state with minimal $g_n$ is the stable state, transitions downward are more likely, but upward transitions still have nonzero rates. The outcome is {\em diffusion} over quasienergy states away from the minimum of $g$ (or the extremum of $g$, for an additively modulated oscillator, see Fig.~\ref{fig:quasienergy_surfaces}), that accompanies {\em drift} (relaxation) toward the minimum (extremum) of $g$.

For a thermal equilibrium system with nondegenerate energy levels $E_{\alpha}$, the ratio of the rates of interstate transitions $|\alpha\rangle\to|\beta\rangle$ and $|\beta\rangle\to|\alpha\rangle$ due to weak coupling to a bath is $W_{\alpha\beta}^{\rm th}/W_{\beta\alpha}^{\rm th}=\exp[(E_{\alpha}-E_{\beta})/k_BT] $ and is fully determined by temperature. Similarly, the ratio $W_{nm}/W_{mn}$ of the transition rates between quasienergy states $|n\rangle\to |m\rangle$ and $|m\rangle \to |n\rangle$ characterizes the effective temperature ${\mathcal T}_e$ of the distribution over these states. It is nonzero even where the bath temperature $T=0$.

A modulated oscillator does not have detailed balance\index{detailed balance} for $T> 0$. The transitions $|n\rangle \to |m\rangle$ are not limited to $m=n\pm 1$ (in which case detailed balance would hold automatically), and the ratio $W_{nm}/W_{mn}$ cannot be written as $\exp[(\ep_n-\ep_m)/{\mathcal T}_e]$ with the same effective temperature ${\mathcal T}_e$ for all $n$ and $m$. In other words, the stationary distribution is generally not of the Boltzmann form, it can be described by an $\ep$- or, equivalently, $g$-dependent temperature.

A complete analysis of the distribution can be done using the master equation\index{master equation} for the oscillator density matrix $\rho$.\index{density matrix} In slow dimensionless time $\tau$, for the coupling to a thermal reservoir (\ref{eq:linear_relaxation}) this equation reads
\begin{eqnarray}
\label{eq:master_eq}
\dot\rho\equiv\partial_{\tau}\rho=&& i\lambda^{-1}[\rho, \hat g]-\hat\kappa\rho, \qquad \hat\kappa\rho= \kappa(\bar n + 1)(\ad a\rho-2a\rho\ad+\rho \ad a)\nonumber\\
&&+\kappa\bar n(a\ad\rho-2\ad\rho a+\rho a\ad), \qquad \kappa = \hbar\Gamma/\lambda E_{\rm sl}.
\end{eqnarray}
Here, the term $\propto [\rho,\hat g]$ describes dissipation-free motion, cf. eqn.~(\ref{eq:Schrodinger_eq}). Operator $\hat \kappa \rho$ describes dissipation and has the same form as in the absence of oscillator modulation \shortcite{Mandel1995,DK_review84}, $\kappa $ is the dimensionless decay rate, see Table~\ref{table:parameters}, $a$ is the lowering operator, and $\bar n$ is the oscillator Planck number,
\begin{equation}
\label{eq:a_interms_Q}
a=(2\lambda)^{-1/2}(Q+iP),\qquad \bar n \equiv \bar n(\omega_0) =[\exp(\hbar\omega_0/k_BT)-1]^{-1}
\end{equation}
[in the lab frame, $a$ has an extra factor $\exp(-i\omega_{\rm M}t)]$; the oscillator eigenfrequency $\omega_0$ is defined so as to incorporate the renormalization due to the coupling (\ref{eq:linear_relaxation}).

\subsection{Semiclassical dynamics condition}
\label{subsec:semiclass}

\index{semiclassical dynamics}Of utmost interest is the parameter range where quantum fluctuations about the states of the forced vibrations of the oscillator are small compared to the interstate distance in phase space and the oscillator motion in the rotating frame is semiclassical. In terms of Fig.~\ref{fig:relaxation_sketch}, it means that the number of states in the wells of $g_{\rm par}(Q,P)$ is large.
Similarly, a large number of states are localized near the extrema of $g_{\rm add}(Q,P)$, for additive modulation. As seen from eqns~(\ref{eq:g_additive}) - (\ref{eq:Schrodinger_eq}), this requires the dimensionless Planck constant to be small,
\begin{equation}
\label{eq:small_lambda}
\lambda \ll 1.
\end{equation}
In the range (\ref{eq:small_lambda}) the rate of switching between the stable states of forced vibrations $W_{\rm sw}$ is exponentially small, $-\ln W_{\rm tun}\propto \lambda^{-1}$ for $T=0$, see Sec.~\ref{sec:quantum_activation}.
We will assume that $W_{\rm sw} \ll \Gamma$. Then over time $\sim \Gamma^{-1}$ the oscillator will reach the state of forced vibrations in the vicinity of which it was prepared initially and will then fluctuate about it; switching between the states takes an exponentially longer time, see Sec.~\ref{sec:quantum_activation}.

\subsection{Quantum temperature}
\label{subsec:quantum_T}

Near the extrema function $g(Q,P)$ is parabolic. The oscillator motion in the rotating frame is mostly harmonic vibrations about these extrema provided the quantum smearing $\propto\lambda^{1/2}$ is small compared to the scale in phase space where the nonparabolicity of $g(Q,P)$ becomes substantial (see also Sec.~\ref{sec:q_heat_spectra}). If we disregard dissipation, the motion is described by the Heisenberg equations
\[\dot Q\equiv dQ/d\tau =-i\lambda^{-1} [Q,\hat g], \qquad \dot P\equiv dP/d\tau = -i\lambda^{-1}[P,\hat g]
\qquad (\Gamma \to 0)
\]
linearized in $Q-Q_0, P$. Here, $Q_0,P=0$ is the position of the considered extremum of $g$. The dimensionless vibration frequency is $\nu_0=\left|g_{QQ}g_{PP}\right|^{1/2}$ [the derivatives of $g(Q,P)$ are calculated for $Q=Q_0,P=0$]. Its dependence on the oscillator parameters is shown in Fig.~\ref{fig:nonequidist_parameters}. In the ground vibrational state the average values of $(Q-Q_0)^2$ and $P^2$ are different, which indicates that forced vibrations in the lab frame are squeezed.

It is convenient to change from $Q-Q_0,P$ to the appropriate raising and lowering operators $b^{\dag}$ and $b$ using the standard squeezing\index{squeezing} transformation
\begin{eqnarray}
\label{eq:squeezed_operators}
&&
Q-Q_0+iP=(2\lambda)^{1/2}(b\cosh \varphi_* - b^{\dag}\sinh \varphi_*),\nonumber\\
&&\hat g\approx g(Q_0,0) + \lambda\nu_0\left(b^{\dag}b+1/2\right){\rm sgn}g_{QQ},
\qquad \nu_0=\left|g_{QQ}g_{PP}\right|^{1/2},
\end{eqnarray}
which makes the mapping on small-amplitude vibrations explicit; in eqn~(\ref{eq:squeezed_operators}), $\tanh \varphi_*=(|g_{QQ}|^{1/2}-|g_{PP}|^{1/2})/(|g_{QQ}|^{1/2}+|g_{PP}|^{1/2})$.

The distribution over quasienergy states near the extrema of $g(Q,P)$ can be found from eqn~(\ref{eq:master_eq}). For small damping, $\kappa\ll \nu_0$, off-diagonal matrix elements of $\rho$ in the basis of quasienergy wave functions are small, $|\rho_{nm}|\ll \rho_{nn}, \rho_{mm}$ for $m\neq n$. To find the diagonal matrix elements one should substitute into eqn~(\ref{eq:master_eq}) $\hat g$ from (\ref{eq:squeezed_operators}) and, using eqns.~(\ref{eq:a_interms_Q}) and (\ref{eq:squeezed_operators}), express $a, \ad$  in $\hat\kappa\rho$ in terms of operators $b,b^{\dag}$ keeping only bilinear terms that contain both $b$ and $b^{\dag}$ while disregarding terms with $b^2,(b^{\dag})^2$ as well as the terms linear in $b,b^{\dag}$. Then  operator $\hat\kappa\rho$ in terms of $b, b^{\dag}$ becomes of the same form as in terms of $a,a^{\dag}$, except that $\bar n$ is replaced with $\bar n_e$,
\begin{equation}
\label{eq:n_e}
\bar n_e=\bar n + (2\bar n + 1)\sinh^2\varphi_*, \qquad{\mathcal T}_e=\lambda\nu_0/ \ln[(\bar{n}_e+1)/\bar{n}_e].
\end{equation}
The stationary solution of the resulting equation is of the Boltzmann type, $\rho^{\rm (st)}=(\bar n_e+1)^{-1} \exp(-\lambda\nu_0b^{\dagger}b/{\mathcal T}_e)$ \shortcite{Dykman2011}. Parameter ${\mathcal T}_e$ given in eqn.~(\ref{eq:n_e}) is the effective dimensionless temperature of vibrations about the stable state in the rotating frame.\index{effective temperature}
For $\bar n=0$ the result coincides with that for a driven oscillator resonantly coupled to a two-level system (Peano and Thorwart \citeyearNP{Peano2010,Peano2010a}). The distributions over quasienergy states for other systems and other relaxation mechanisms were discussed by \shortciteN{Verso2010} and \shortciteN{Ketzmerick2010}.
\begin{figure}[t]
\includegraphics[width=4.0 cm]{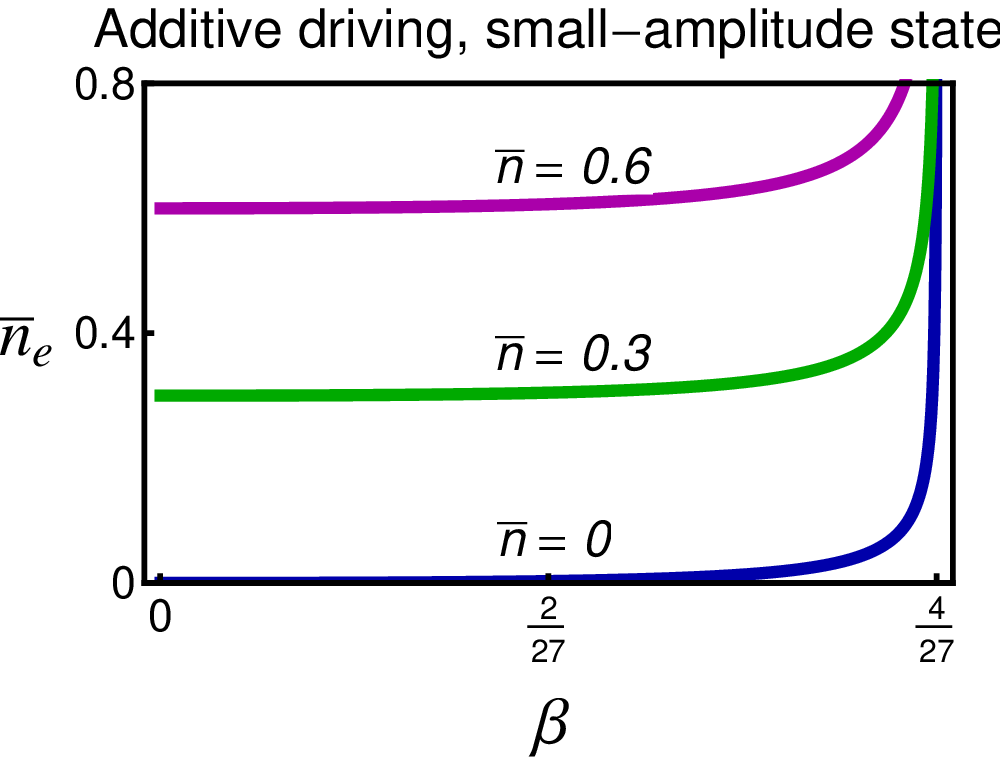}\hfill
\includegraphics[width=3.8 cm]{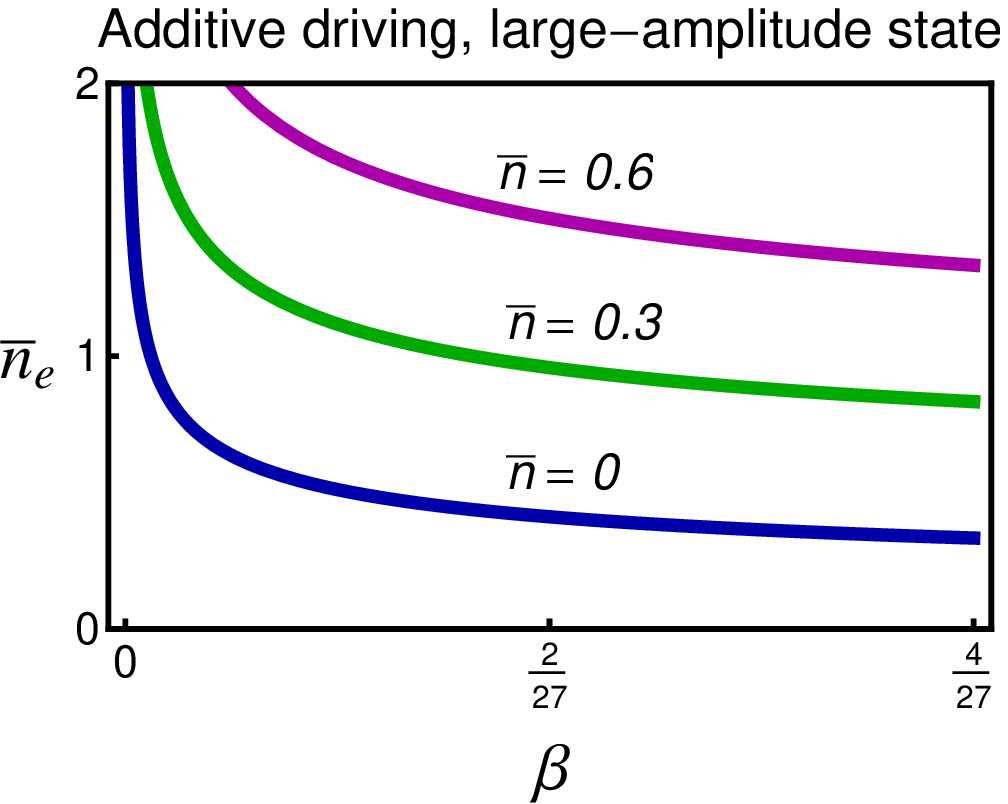}\hfill
\includegraphics[width=4.1 cm]{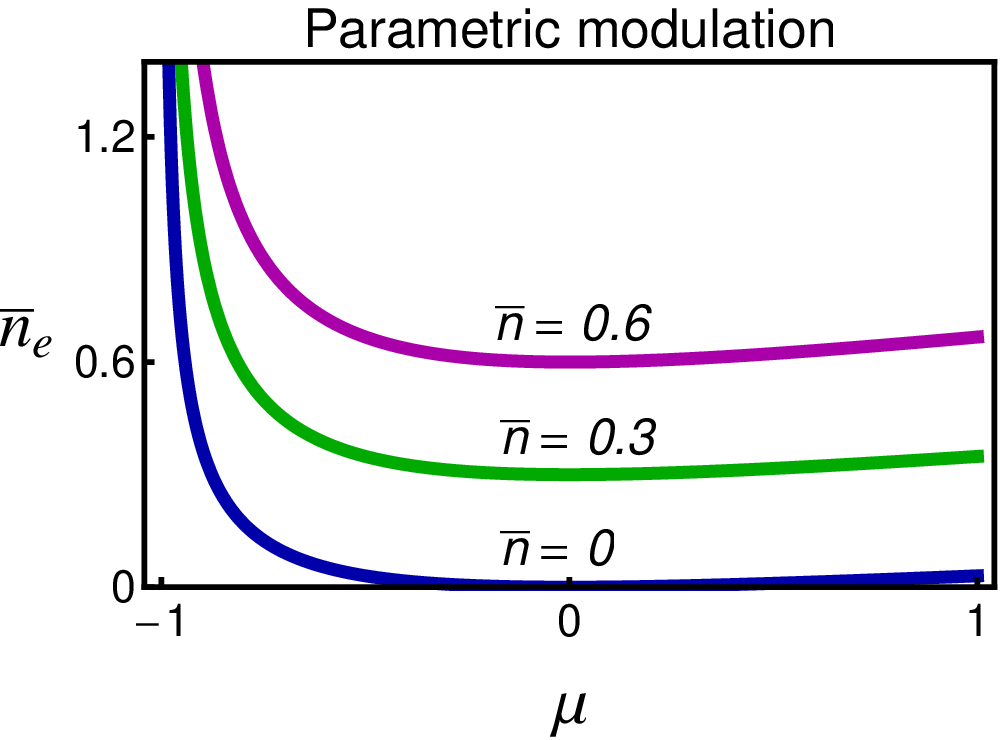}
\caption{The effective Planck number $\bar n_e$ for vibrations about the state of forced vibrations. The left, central, and right panels refer, respectively, to the small- and large-amplitude states of an additively modulated oscillator and to a parametrically modulated oscillator. The damping is assumed to be small, $\kappa\ll \nu_0$ in eqn~(\ref{eq:master_eq}).}
\label{fig:T_eff}
\end{figure}
%

From eqn~(\ref{eq:n_e}), $\bar n_e=\sinh^2\varphi_* > 0$ even where the Planck number of the original oscillator is zero. In the opposite limit of high temperature, $k_BT\gg \hbar\omega_0$, we have ${\mathcal T}_e\propto T$, and $\rho^{\rm (st)}$ looks like a Boltzmann distribution of an oscillator with frequency $\omega_0/(1+2\sinh^2\varphi_*) < \omega_0$. The dependence of $\bar n_e$ on the  parameter that characterizes a resonantly modulated oscillator in the small damping limit is shown in Fig.~\ref{fig:T_eff}.

Expression (\ref{eq:n_e}) is simplified also near the bifurcation point \index{bifurcation!point}(the bifurcation parameter value) where the corresponding stable vibrational state disappears. If dissipation is disregarded, at the bifurcation point the corresponding extremum (two extrema, for parametric oscillator) and the saddle point of $g(Q,P)$ merge, whereas for the large-amplitude state for additive modulation, the values of $g(Q,P)$ at the extremum and the saddle point coincide. As the parameters $\beta$ or $\mu$ approach their bifurcation values $\beta_B$ or $\mu_B$, see Fig.~\ref{fig:bistability},  $g_{QQ}\to 0$ or, for the large-amplitude state of an additively modulated oscillator ($\beta\to\beta_{B1}$), $g_{PP}\to 0$. Near a bifurcation point $\nu_0\ll 1$. Then ${\mathcal T}_e\approx (\lambda/4)(2\bar n+1)|g_{PP}|$; for the large-amplitude state of an additively modulated oscillator $g_{PP}$ should be replaced with $g_{QQ}$. As we see, ${\mathcal T}_e$ displays a characteristic temperature dependence described by the factor $2\bar n+1$. To the leading order, it is independent of the distance to the bifurcation point $\eta=\beta-\beta_B$ or $\eta=\mu-\mu_B$. In contrast, $\bar n_e\propto \nu_0^{-1}$ sharply increases with decreasing $|\eta|$. This is because the system becomes ``soft" near a bifurcation point, and respectively, the distribution broadens.  The scaling of $\bar n_e$ with $\eta$ is $\bar n_e\propto\nu_0^{-1}\propto |\eta|^{-\xi_T}$ with $\xi_T=1/2$ for parametric modulation\index{modulation!parametric} and $\xi_T=1/4$ for additive modulation.\index{modulation!additive}

The results for ${\mathcal T}_e$ apply for not too small $|\eta|$. One constraint on $|\eta|$ is classical, $\nu_0\gg \kappa$. The other is quantum: the dimensionless spacing of quasienergy levels $\lambda\nu_0\propto |\eta|^{\xi_T}$ should be small compared to the difference $\Delta g$ between the values of $g$ at the extremum and the saddle point of $g(Q,P)$ that merge at the bifurcation point; for the corresponding extremum, for parametric modulation $\Delta g\propto \eta^2$, whereas for additive driving $\Delta g\propto \eta^{3/2}$ for $\beta\to\beta_{B2}$ and $\Delta g\propto \eta^{1/2}$ for $\beta\to\beta_{B1}$.

%
\section{Quantum activation}
\label{sec:quantum_activation}

\index{quantum activation}The diffusion over quasienergy states, that underlies quantum heating, also populates states near the top of the quasienergy barrier in Fig.~\ref{fig:relaxation_sketch}(b). As a result, if the system was initially occupying the left well of $g(Q,P)$, for example, it will diffuse to the barrier top and switch to the right well. Such an overbarrier transition reminds the conventional switching via thermal activation in systems close to thermal equilibrium \shortcite{Kramers1940}, where the states near the barrier top are populated as a result of thermal fluctuations. In the case of a modulated oscillator, for low temperature the effect is due to quantum fluctuations. Respectively, the switching mechanism can be called quantum activation. It applies to both parametrically and additively modulated oscillators: in the latter case switching occurs, with probability $\sim 1/2$, once the oscillator located initially near an extremum of $g(Q,P)$ reaches the saddle, see Fig.~\ref{fig:quasienergy_surfaces}.\index{switching}

For small effective Planck constant $\lambda$, the switching rate\index{switching!rate} is exponentially small,
\begin{equation}
\label{eq:define_R_A}
W_{\rm sw}\propto\exp(-R_A/\lambda).
\end{equation}
This estimate can be easily understood from Fig.~\ref{fig:relaxation_sketch}(b). If the ratio of the typical rates of transitions up and down in quasienergy is $W_{\uparrow}/W_{\downarrow}<1$ and the transitions occur primarily between a few neighboring quasienergy states,
the population of the levels close to the barrier top is $\sim (W_{\uparrow}/W_{\downarrow})^M$, where $M$ is the number of intrawell states. Since the dimensionless level spacing $g_{n+1}-g_{n}\sim \lambda$ and the well depth $g_{\max}-g_{\min} \sim 1$, we have $M\sim \lambda^{-1}$, which immediately gives $-\ln W_{\rm sw}\sim \lambda^{-1}$ for low temperatures. This estimate applies to both additively and parametrically modulated oscillators.

Even though the effective activation energy\index{effective activation energy} for a given stable state $R_A$ is determined by quantum fluctuations for low $T$, quantum activation is {\em not} tunneling, it is the result of coupling of the oscillator to a thermal bath and the quantum noise that accompanies relaxation. Therefore finding $R_A$ requires solving the master equation (\ref{eq:master_eq}). For small $\lambda$ this can be done using the WKB approximation. The problem is qualitatively different from that of switching in systems close to thermal equilibrium, which can be efficiently approached using the instanton technique \cite{Langer1967,Coleman1977,Affleck1981,Caldeira1983}. A modulated oscillator does not have detailed balance, generally, and its distribution is not of the Boltzmann form and is not characterized by the partition function.

For small relaxation rate, switching of a modulated oscillator can also occur via dynamical tunneling \index{tunneling!dynamical}with constant quasienergy, for example via tunneling between equal-quasienergy states in the left and right wells of $g(Q,P)$ in Fig.~\ref{fig:relaxation_sketch}(b), see \shortciteNP{Larsen1976,Sazonov1976,Dmitriev1986,Vogel1988,Wielinga1993,Peano2006,Marthaler2007a,Serban2007}. As we explain below, switching via tunneling becomes substantial only where the relaxation rate of the oscillator is exponentially small, otherwise switching occurs via quantum activation.

\subsection{The WKB switching paths}
\label{subsec:optimal_path}
The rate of switching from a given vibrational state can be found from the quasi-stationary solution of the master equation. It is formed on times $\Gamma^{-1}\ll t\ll W_{\rm sw}^{-1}$ in a broad range of phase space. As mentioned previously, in this time domain, the oscillator prepared at $t=0$ near the considered stable state in phase space will have come to local equilibrium, but most likely will not have switched to another state.

\begin{figure}[h]
\begin{center}
\includegraphics[width=4.0 cm]{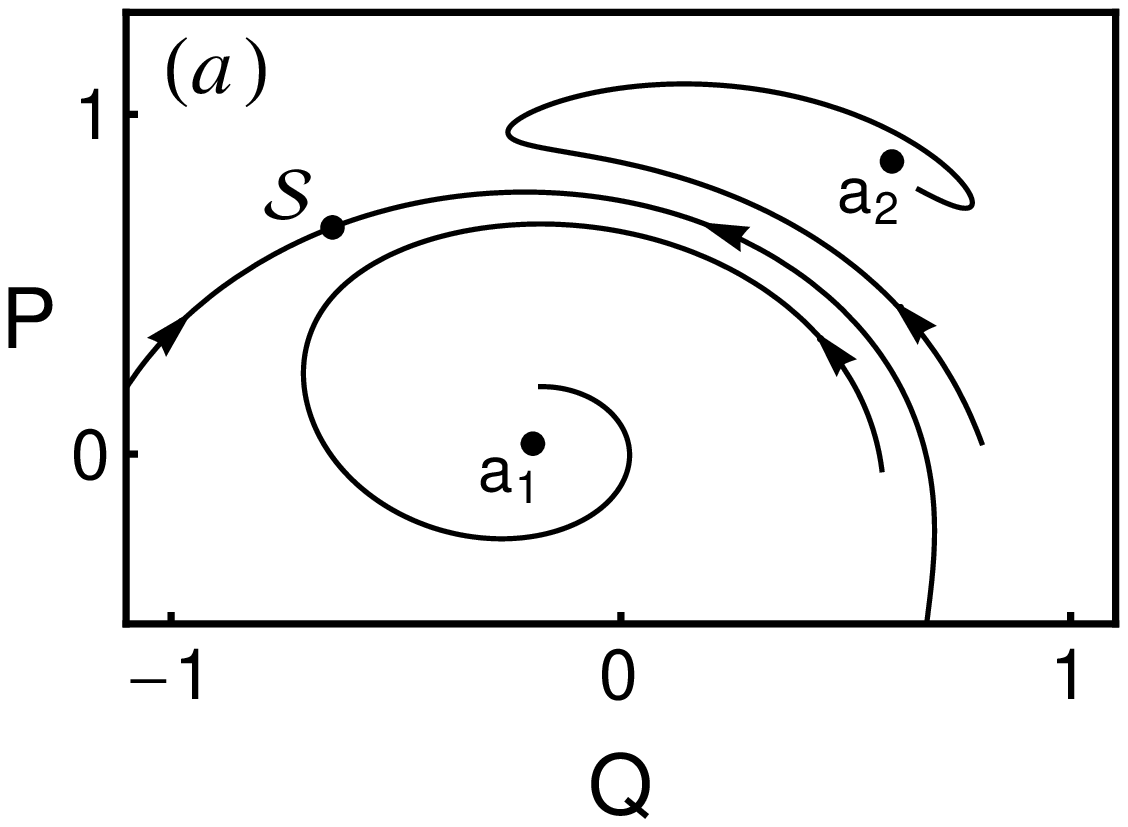}\hspace*{1.0in}\hspace{0.1in}
\includegraphics[width=4.0 cm]{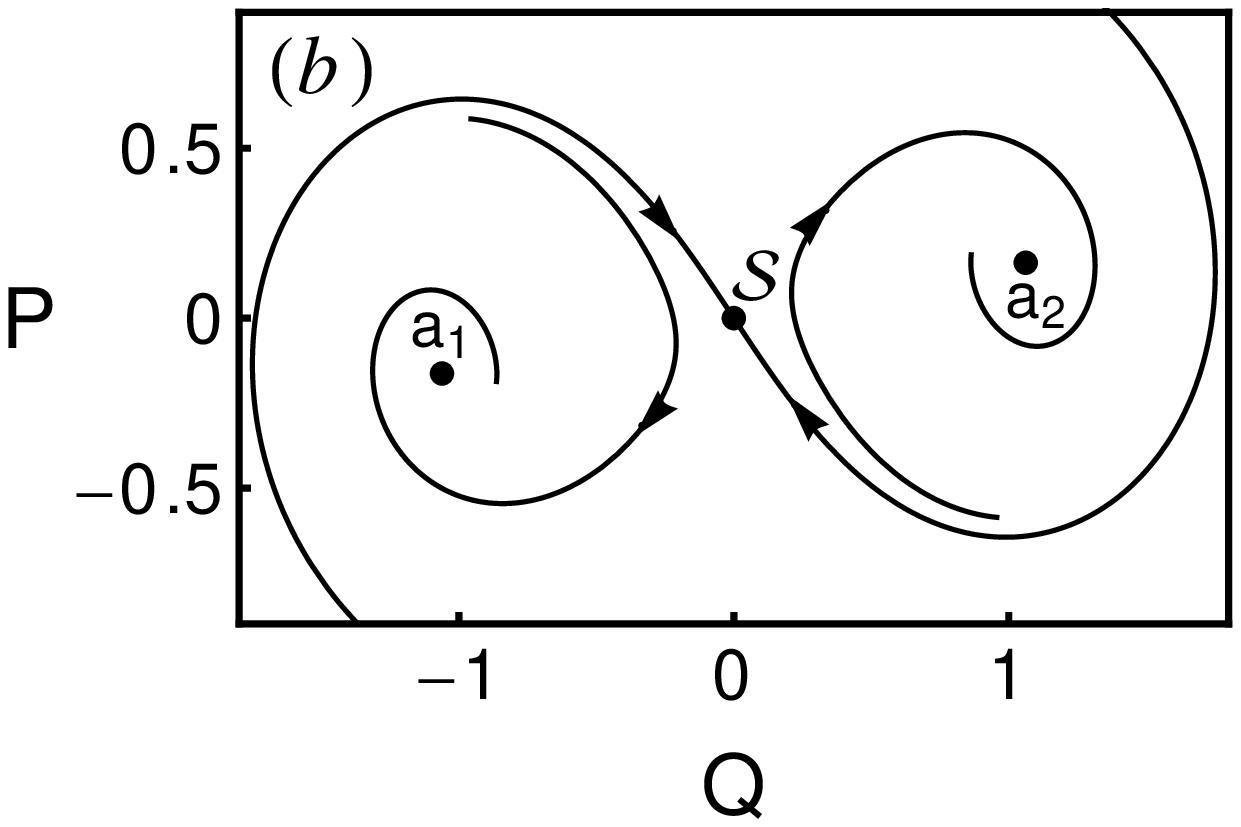}
\end{center}
\caption{The phase portraits of the classical additively (a) and parametrically (b) modulated oscillators in the rotating frame. The attractors a$_{1,2}$ correspond to the stable states of forced vibrations in the lab frame. Their positions $(Q_{\rm a},P_{\rm a})$ give the  scaled vibration amplitudes $A_{\rm osc}= (Q_{\rm a}^2+P_{\rm a}^2)^{1/2}$ in Fig.~\protect\ref{fig:bistability}. The separatrix that goes through the saddle point $\mathcal S$ is the boundary of the basins of attraction to different attractors. The phase portrait in (b) has inversion symmetry. The parameters in (a) are $\beta=1/27,\kappa=0.15$; in (b) $\mu=0.2,\kappa=0.3$.}
\label{fig:phase_portrait}
\end{figure}

The physical picture of switching can be understood from the classical phase portrait\index{phase portrait} for the range of bistability \index{bistability}shown in Fig.~\ref{fig:phase_portrait}. The positions of the stable states on the oscillator phase plane in the rotating frame $(Q_{\rm a},P_{\rm a})$ are given by the stable solutions of classical equations of motion
\begin{equation}
\label{eq:classical_eom}
\dot Q =\partial_Pg(Q,P)-\kappa Q,\qquad \dot P =-\partial_Qg(Q,P)-\kappa P.
\end{equation}
These equations immediately follow from master equation (\ref{eq:master_eq}) if in the equations for the average values $\langle Q\rangle = {\rm Tr}~Q\rho, \langle P\rangle = {\rm Tr}~P\rho$ one disregards fluctuations, $\langle Q^nP^m\rangle \to \langle Q\rangle^n\langle P\rangle^m\to Q^nP^m$.

For $t\ll W_{\rm sw}^{-1}$, the oscillator is mostly localized in phase space near the initially occupied stable state\index{stable state}, with localization length $\sim \lambda^{1/2}$ for low temperatures. The tail of its quasi-stationary distribution is formed by large rare fluctuations. In the semiclassical picture, switching occurs as a result of a large fluctuation that brings the oscillator to the basin of attraction of the initially empty stable state. From there, the oscillator  will most likely move to this state staying within $\propto \lambda^{1/2}$ from the classical trajectory (\ref{eq:classical_eom}). The rate $W_{\rm sw}$ is determined by the maximal rate of an appropriate fluctuation.

In this section we will be interested in finding the switching exponent, i.e., the leading-order term in $\ln W_{\rm sw}$. This can be done \shortcite{Dykman1988a} using  the density matrix in the coordinate representation $\rho(Q_1,Q_2)\equiv \langle Q_1|\rho|Q_2\rangle$.\index{density matrix!coordinate representation} Equation~(\ref{eq:master_eq}) in this representation has the form
\[\dot\rho(Q_1,Q_2)=-i\lambda^{-1}{\mathcal H}(Q_1,Q_2,-i\lambda\partial_{Q_1},-i\lambda\partial_{Q_2})\rho(Q_1,Q_2),\]
where
\begin{eqnarray}
\label{eq:two_particle_H}
{\mathcal H}(Q_1,Q_2,P_1,P_2) &=& g(Q_1,P_1)-g(Q_2,P_2) -\kappa(P_1Q_2 + P_2Q_1-i\lambda)\nonumber\\
&&-\frac{1}{2}i\kappa(2\bar n+1)\left[ (Q_1 - Q_2)^2 + (P_1+P_2)^2\right].
\end{eqnarray}

Along with $\rho(Q_1,Q_2)$ it is convenient to consider the density matrix in the Wigner representation\index{density matrix!Wigner representation}
\begin{equation}
\label{eq:wigner}
\rho_W(Q,P)=\int d\xi e^{-i\xi P/\lambda}\rho(Q+\xi/2,Q-\xi/2).
\end{equation}
For the oscillator, function $\rho_W$ in the quasi-stationary regime has a Gaussian peak at the initially occupied stable state $(Q_{\rm a},P_{\rm a})$; this form of $\rho_W$ is generic for semiclassical systems (for $\kappa\ll \nu_0$ it follows from the results of Sec.~\ref{subsec:quantum_T}). It rapidly falls off away from $Q_{\rm a},P_{\rm a}$. The switching exponent is determined by the maximal value of $\rho_W(Q,P)$ for $Q,P$  inside the basin of attraction of the initially empty stable state or on the basin boundary, i.e., it is determined by the tail of $\rho_W(Q,P)$ and in turn, by the tail of $\rho(Q_1,Q_2)$.

In the spirit of the WKB approximation\index{WKB approximation}, one can seek $\rho(Q_1,Q_2)$ on the tail in the eikonal form.\index{eikonal form} To the leading order in $\lambda$ in the quasi-stationary regime ($\dot \rho = 0$) we have
\begin{equation}
\label{eq:eikonal_Q}
\rho(Q_1,Q_2)=\exp[iS(Q_1,Q_2)/\lambda], \qquad{\mathcal H}(Q_1,Q_2,\partial_{Q_1}S,\partial_{Q_2}S)=0.
\end{equation}
Equation (\ref{eq:eikonal_Q}) has the form of the Hamilton-Jacobi equation for an auxiliary {\it classical} system with coordinates $Q_1,Q_2$ and action $S$, with equations of motion
\begin{equation}
\label{eq:Hamilton_trajectories}
\dot Q_j=\partial_{P_j}{\mathcal H}(Q_1,Q_2,P_1,P_2),\qquad \dot P_j=-\partial_{Q_j}{\mathcal H}(Q_1,Q_2,P_1,P_2)\qquad (j=1,2).
\end{equation}

Equations~(\ref{eq:eikonal_Q}) and (\ref{eq:Hamilton_trajectories}) map the problem of finding the tail of the density matrix of a dissipative quantum oscillator onto the problem of finding classical Hamiltonian trajectories. The initial conditions for these trajectories follow from the Gaussian form of $\rho_W$ near $Q_{\rm a},P_{\rm a}$. From (\ref{eq:wigner}) and (\ref{eq:eikonal_Q}), for the trajectories coming from the stable state of the oscillator the initial conditions are $Q_1=Q_2=Q_{\rm a}$ and $P_1=-P_2=P_{\rm a}$, where $P_{1,2}=\partial_{Q_{1,2}}S$. The analysis shows that the final point on the switching trajectory (\ref{eq:Hamilton_trajectories}) is $Q_1=Q_2=\QS$ and $P_1=-P_2=\PS$, where $(\QS,\PS)$ is the saddle point of the oscillator.

The effective activation energy for switching from a given stable state is determined by action $S$ calculated along the switching trajectory,
\begin{equation}
\label{eq:R_A_from_trajectory}
R_A={\rm Im}\int\nolimits_{-\infty}^{\infty}d\tau \sum\nolimits_{j=1,2}P_j\dot Q_j.
\end{equation}
We took into account that the initial and final points on the trajectory are stationary states of the auxiliary system, therefore the time integral goes from $-\infty$ to $\infty$.

The formulation (\ref{eq:two_particle_H}) - (\ref{eq:R_A_from_trajectory}) reminds the conventional instanton formulation. The major distinction is that the motion occurs in {\em real} rather than imaginary time. At the same time, the switching trajectory is {\em complex}, as are also ${\mathcal H}$ and $S$. It is instructive to compare this formulation with the theory of noise-induced switching in classical systems. There, for Gaussian noise, the switching rate displays activation dependence on the noise intensity ($k_BT$, for thermal noise). The effective activation energy can be calculated as action of an auxiliary Hamiltonian system \shortcite{Freidlin_book,Dykman1990,Kamenev2011}. The trajectory followed by the auxiliary system is real and gives the most probable trajectory that the initial noise-driven system follows in switching. Such trajectory has been seen in experiment, see the chapter by H. B. Chan and C. Stambaugh in this book and references therein.

One can see that, due to the symmetry of the Hamiltonian ${\mathcal H}$, eqn~(\ref{eq:Hamilton_trajectories}) has a solution $Q_2(t)=Q_1^*(t), P_2(t)=-P_1^*(t)$, which satisfies the boundary conditions. This solution was used to find the activation energy $R_A$ for an additively modulated oscillator as function of $\beta$ for small damping \shortcite{Dykman1988a}.

\subsection{Balance equation approach}
\label{subsec:balance_equation}

\index{balance equation}We now discuss an alternative approach, which immediately gives the switching rate and the distribution of the oscillator in the small damping limit. If the broadening of the quasienergy levels is small compared to the interlevel distance, one can disregard off-diagonal matrix elements $\rho_{nm}$ $(n\neq m)$ in the basis of quasienergy wave functions. Then eqn~(\ref{eq:master_eq}) is reduced to a balance equation for state populations
\begin{equation}
\label{eq:balance_eqn}
\dot\rho_{nn}=\sum\nolimits_m\left(W_{mn}\rho_{mm}-W_{nm}\rho_{nn}\right), \quad W_{mn}=2\kappa\left[(\bar n + 1)|a_{nm}|^2 +\bar n |a_{mn}|^2\right],
\end{equation}
where $a_{nm}\equiv \langle n|a|m\rangle$ (we remind that $a$ is the oscillator lowering operator). We disregard tunneling when defining functions $|n\rangle\equiv \psi_n(Q)$, i.e., we use the ``intrawell" wave functions in Fig.~\ref{fig:relaxation_sketch}(b); the effect of tunneling is exponentially small for $\lambda \ll 1$.

Matrix elements $a_{mn}$ can be calculated in an explicit form using the WKB approximation. It relates the problem to that of classical conservative motion $\dot Q=\partial_Pg, \dot P=-\partial_Qg$. Such motion is periodic oscillations in time $Q(\tau;g),P(\tau;g)$ with given $g(Q,P)=g$ and with dimensionless frequency $\nu(g)$ that depends on $g$. For not too large $|m-n|$, matrix element $a_{mn}$ is given by the $(m-n)$th Fourier component of the periodic function $a(\tau;g_n)=(2\lambda)^{-1/2}\left[Q(\tau;g)+iP(\tau;g)\right]$ calculated for the classical orbit $g(Q,P)=g_n$. Formally, we require that $|m-n|\ll n$, but the results apply also near the extrema of $g(Q,P)$ where $n\sim 1$, since $\psi_n(Q)$ are close to the wave functions of a harmonic oscillator for such $n$.

The evaluation of $a_{mn}$ simplifies if one notices that $g(Q,P)$ is quartic in $Q,P$ and even in $P$. Because of that, the orbits $Q(\tau;g), P(\tau;g)$ are described by the Jacobi elliptic functions and are double periodic in $\tau$. To calculate $a_{mn}$ one can then integrate $a(\tau;g_n)\exp[i\tau (n-m)\nu(g_n)]$ along an appropriately chosen closed contour on the complex $\tau$-plane. The result is determined by the pole of $a(\tau;g_n)$ and has a simple form \shortcite{Marthaler2006}. In particular, $W_{mn}$ exponentially decays with $|m-n|$ for $1\ll |m-n|$, with the exponent that depends on the sign of $m-n$.

The quasi-stationary populations of neighboring states $n$ and $n\pm 1$ in eqn~(\ref{eq:balance_eqn}) can differ significantly. However, as we will see, $\ln\rho_{nn}$ is a smooth function of $n$. Respectively, we seek the quasi-stationary distribution in the eikonal form,\index{eikonal form} $\rho_{nn}=\exp(-R_n/\lambda)$, $R_n\equiv R(g_n)$. To the leading order in $\lambda$, eqn~(\ref{eq:balance_eqn}) then reads
\begin{eqnarray}
\label{eq:eikonal_R_prime}
\sum\nolimits_k W_{n+k\,n}\left\{1-\exp\left[-k\nu(g_n)R'(g_n)\right]\right\}=0,\qquad R'(g)\equiv dR/dg,
\end{eqnarray}
where we used $g_{n+k}\approx g_n+\lambda k\nu(g_n),\,R_{n+k}\approx R_{n} + \lambda k\nu(g_n)R'(g_n)$ and $W_{n\,n-k}\approx W_{n+k\,n}$ for $|k|\ll n$.

From eqn~(\ref{eq:eikonal_R_prime}), $R'(g)$ is independent of $\lambda$ and is given by a solution of a polynomial equation. For $g$ close to its value $g_{\rm a}=g(Q_{\rm a},P_{\rm a})$ at a stable state  ($P_{\rm a}\to 0$ for $\kappa\to 0$), the solution of eqn~(\ref{eq:eikonal_R_prime}) is of the Boltzmann form, $\rho_{nn}\propto \exp[-\lambda\nu_0nR'(g_{\rm a})]$; as one can show, it coincides with the result of Sec.~\ref{subsec:quantum_T}, with $R'(g_{\rm a})=\lambda/{\mathcal T}_e$. However, this is only the asymptotic solution, generally the distribution is not described by an effective temperature, because $R'(g)$ varies with $g$. We note that the corrections disregarded in deriving eqn~(\ref{eq:eikonal_R_prime}) are $\propto\lambda$, which justifies this equation for $\lambda\ll 1$.

\begin{figure}[h]
\begin{center}
\includegraphics[width=4.0 cm]{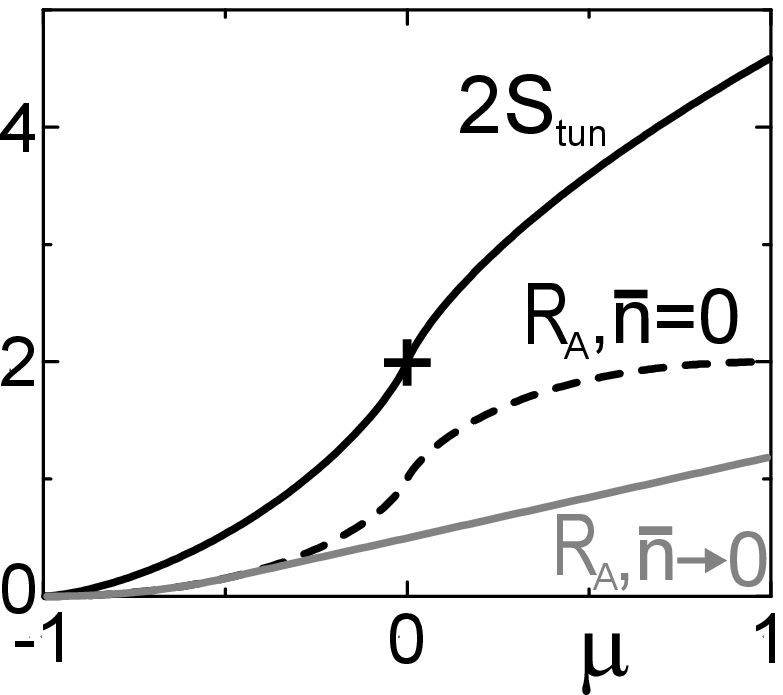}\hspace*{1.0in}
\includegraphics[width=4.4 cm]{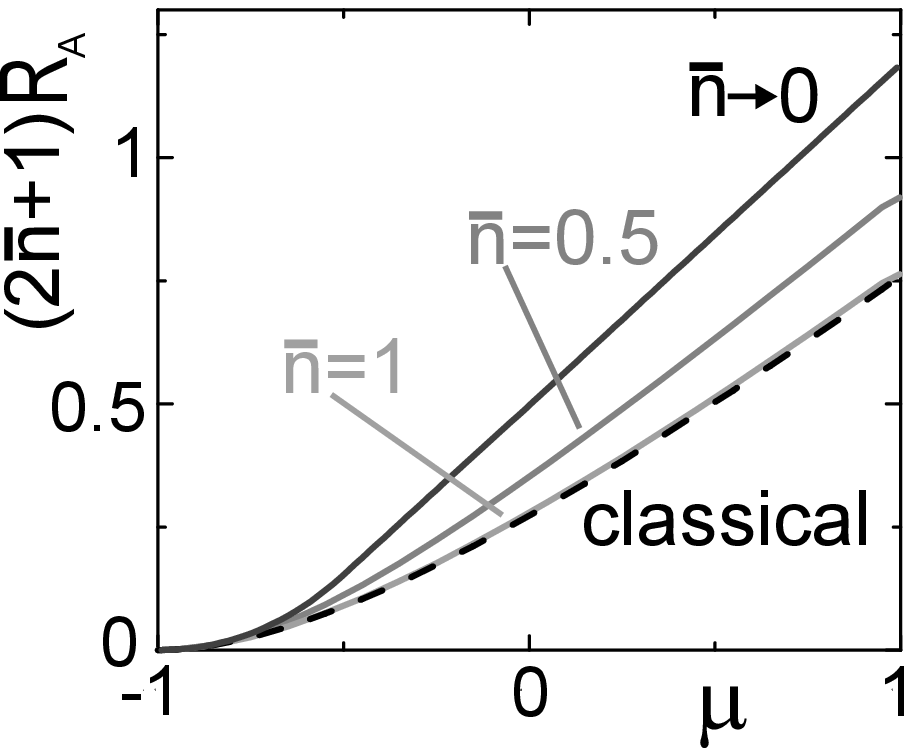}
\end{center}
\caption{The tunneling exponent $S_{\rm tun}$ for tunneling between the extrema of $g(Q,P)$ of a parametric oscillator and the switching activation energy $R_A$ in the limit of small damping for different values of the Planck number \protect\shortcite{Marthaler2006}. The cross indicates the value of $S_{\rm tun}$ obtained by \protect\shortciteN{Wielinga1993}.}
\label{fig:q_activ_parametric}
\end{figure}

The effective activation energy\index{effective activation energy} of switching from a given stable state is determined by the quasi-stationary occupation of states near the saddle point, with dimensionless quasienergy $g_{\mathcal S}\equiv g(\QS,\PS)$. Therefore, to the leading order in $\lambda$
\begin{equation}
\label{eq:R_A_via_R_prime}
R_A= \int\nolimits_{g_{\rm a}}^{g_{\mathcal S}}dg R'(g).
\end{equation}
In Fig.~\ref{fig:q_activ_parametric} we show the activation energy of switching between period-two states of a parametric oscillator obtained from eqns~(\ref{eq:eikonal_R_prime}) and (\ref{eq:R_A_via_R_prime}) \shortcite{Marthaler2006}. For small damping, $R_A$ depends on two parameters, the scaled frequency detuning of the modulating field $\mu$ and the Planck number of the oscillator $\bar n$. As seen in the right panel, the value of $(2\bar n + 1)R_A$ decreases with increasing temperature and already for $\bar n=1$ becomes very close to the result for the classical range $\bar n\gg 1$. In this range, switching is thermally activated and $R_A\propto 1/T$ \shortcite{Dykman1998}; cf. the experiments by \shortciteN{Lapidus1999} and \shortciteN{Chan2007}. Both in the classical limit and near bifurcation points $|\nu(g)R'(g)|\ll 1$ and eqn~(\ref{eq:balance_eqn}) becomes a linear equation for $R'(g)$.

Near the bifurcation point\index{bifurcation!point} $\mu_{B1}$ in Fig.~\ref{fig:bistability} where the stable period two states merge together and disappear, $R_A\propto (2\bar n +1)^{-1}(\mu-\mu_{B1})^2$. For additive driving, for switching from the large and small-amplitude vibrational states in Fig.~\ref{fig:bistability} near the corresponding bifurcation points\index{bifurcation!scaling} $R_A\propto (2\bar n +1)^{-1}(\beta-\beta_{B1})$ and $R_A\propto (2\bar n +1)^{-1}(\beta_{B2}-\beta)^{3/2}$, respectively \shortcite{Dykman1988a}. This scaling applies not too close to bifurcation points, where still $\nu_0\equiv \nu(g_{\rm a})\gg \kappa$.


\subsubsection{Detailed balance for $T=0$}
\label{subsubsec:detailed_balance}
\index{detailed balance}An important feature of the semiclassical matrix elements, which follows from the double-periodicity of $a(\tau;g)$, is that $|a_{mn}/a_{nm}|^2=\exp[(m-n)c(g_n)]$ for $|m-n|\ll n$, with $c(g_n)$ that smoothly depends on $n$. Therefore for $T=0$ the system has detailed balance: the ratio of the transition rates is path-independent, $W_{nm}W_{mk}/(W_{km}W_{mn})=W_{nk}/W_{kn}$, and eqn~(\ref{eq:balance_eqn}) has a quasi-stationary solution $\rho_{mm}/\rho_{nn}=W_{nm}/W_{mn}$. This gives the quasi-stationary population of states near the saddle point $(\QS,\PS)$ relative to that near the initially occupied stable state $(Q_{\rm a},P_{\rm a})$, and thus the activation energy $R_A=-\int\nolimits_{g_{\rm a}}^{g_{\mathcal S}}dg \nu^{-1}(g)c(g)$. For a parametric oscillator $R_A$ for $T=0$ is shown in Fig.~\ref{fig:q_activ_parametric}.

A remarkable property of the detailed balance solution seen from Fig.~\ref{fig:q_activ_parametric} is fragility: the value of $R_A$ for $T=0$ differs from $R_A$ for $T\to 0$. The fragility emerges for small $\lambda$ in the limit where the relaxation rate $\Gamma\propto \kappa\to 0$. For nonzero $\kappa$, the transition from $T=0$ to nonzero $T$ solutions should be continuous \shortcite{Dykman1988a}; more work is required to study this transition.
The detailed balance condition for $T=0$ applies for arbitrary $\kappa$. This was used to find the stationary probability distribution for both additively and parametrically modulated oscillators  \shortcite{Drummond1980c,Kryuchkyan1996} and the switching rate in the overdamped limit for parametric modulation \shortcite{Drummond1989}.

We now compare switching via quantum activation and dynamical tunneling. The rate of switching via tunneling is $\propto \exp(-2S_{\rm tun}/\lambda)$, where $S_{\rm tun}$ is the tunneling action near the corresponding extremum of $g(Q,P)$. The tunneling rate prefactor is $\propto\nu_0\lambda E_{\rm sl}/\hbar$. The prefactor in the rate of switching via quantum activation for small damping is $\propto \Gamma\propto \kappa$. From Fig.~\ref{fig:q_activ_parametric} and from similar results for additively modulated oscillator, $2S_{\rm tun} > R_A$ \shortcite{Dykman1988a,Marthaler2006}. Therefore unless $\kappa$ is exponentially small, oscillator switches via quantum activation rather than tunneling.


\subsection{Switching rates near bifurcation points}
\label{subsec:bifurcation}

\index{bifurcation}\index{bifurcation!point}The analysis of switching near bifurcation parameter values is particularly important. In this range the switching rates display universal, model-independent features; also, the range is interesting for many applications, in particular, for the Josephson bifurcation amplifiers\index{Josephson!bifurcation amplifier} \shortcite{Vijay2009}. Since the frequency of vibrations about a metastable state $\nu_0$ rapidly decreases with the decreasing distance to a bifurcation point $\eta$, see Sec.~\ref{subsec:quantum_T}, for small $\eta$ the oscillator motion in the rotating frame is often overdamped, $\nu_0\ll \kappa$, which we will assume in this subsection to be the case.

\begin{figure}[t]
\begin{center}
\includegraphics[scale=0.5]{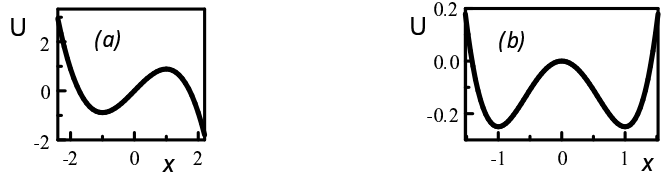}
\end{center}
\caption{Effective potentials for overdamped motion near (a) the saddle-node and (b) the supercritical pitchfork (onset of stable period-two vibrations) bifurcations. The equation of motion in scaled variables is $\dot x = -U'(x)$ with $U(x)=-x^3/3 + \eta x$ in (a) and $U(x)=x^4/4-\eta x^2/2$ in (b); the plots refer to $\eta = 1$. The minimum and maximum of $U$ in (a) correspond to stable and unstable states of forced vibrations for additive modulation; the minima in (b) correspond to the stable vibrations with opposite phase, for parametric modulation. The specific form of $U(x)$ is obtained by keeping the lowest order terms in $x$ and $\eta$ compatible with the condition of merging of one or two stable and an unstable state of the system for $\eta = 0$.}
\label{fig:bif_potentials}
\end{figure}

Near a bifurcation point, the behavior of the system is controlled by a ``soft mode",\index{soft mode} a dynamical variable that slowly changes in time. Without fluctuations the occurrence of such variable is well-known in classical dynamics \shortcite{Guckenheimer1987}. It emerges because, for small $\eta$,  the stable and unstable states of the system are close to each other in phase space. In Fig.~\ref{fig:phase_portrait}(a) the attractor corresponding to stable forced vibrations with large or small amplitude becomes close to the saddle point (the saddle-node bifurcation),\index{bifurcation!saddle-node} whereas in Fig.~\ref{fig:phase_portrait}(b) for small $\mu-\mu_{B1}$ the attractors become close to each other and to the saddle point between them (the supercritical pitchfork bifurcation).\index{bifurcation!pitchfork} In appropriately scaled variables, slow motion along the direction between the close states is described by equation $\dot x = -U'(x)$. The potentials $U(x)$ for the bifurcations of interest are sketched in Fig.~\ref{fig:bif_potentials}.

A theory of the switching rate based on the master equation in the Wigner representation was developed earlier \shortcite{Dykman2007}. Here we sketch a somewhat simpler derivation based on the quantum Langevin equation. Even without writing this equation one can see that, since the motion is overdamped and is characterized by one slow dynamical variable with no conjugate variable, commutation relations for this slow variable are irrelevant. Then its fluctuations are the same as in the case of a classical oscillator \shortcite{Dykman1980}, the only difference being that the fluctuation intensity is determined by quantum rather than classical noise from the thermal bath. From Fig.~\ref{fig:relaxation_sketch}, for nonzero bath temperature the noise intensity is proportional to the overall rate of the bath-induced transitions up and down between the Fock states of the oscillator, which in turn is proportional to $2\bar n + 1$ (the Einstein relation). Respectively, using the quantum to classical correspondence for high temperature, in the expression for the rate of classical thermally activated switching one should replace $k_BT$ with $(2\bar n + 1)\hbar\omega_0/2$, which indeed gives the right answer.

The quantum Langevin equation\index{quantum Langevin equation} in the rotating frame is an extension of the Heisenberg equation for $Q,P$ that includes the effect of coupling to a thermal bath. In the same approximation that led to the Markov master equation (\ref{eq:master_eq}), in slow time $\tau$
\begin{equation}
\label{eq:q_Langevin}
\dot Q=-i\lambda^{-1}[Q,\hat g]-\kappa Q+ \hat f_Q(\tau),\qquad
\dot P=-i\lambda^{-1}[P,\hat g]-\kappa P + \hat f_P(\tau).
\end{equation}
Here, $\hat f_{Q,P}$ are quantum noise operators. Equation~(\ref{eq:q_Langevin}) is well known \shortcite{Ford1965} for a harmonic oscillator linearly coupled to a bath of harmonic oscillators. It applies also for a more general form of the coupling to the bath $H_i$, eqn~(\ref{eq:linear_relaxation}). To the leading order in $H_i$ it can be obtained just by iterating the Heisenberg equations of motion for the bath \shortcite{Lax1966b}. The nonlinearity of our oscillator is relatively weak and does not affect the form of the dissipative and noise terms in eqn~(\ref{eq:q_Langevin}).

From eqn~(\ref{eq:linear_relaxation}), $\hat f_Q,\hat f_P$ are linear combinations of operators $h_{\rm b}(t)\exp(-i\omega_{\rm M}t)$, $h_{\rm b}^{\dag}(t)\exp(i\omega_{\rm M}t)$ calculated disregarding the coupling to the oscillator. For a smooth around $\omega_{\rm M}$ power spectrum of $h_{\rm b}$, the noise is $\delta$-correlated in slow time,
\begin{equation}
\label{eq:noise_correlators}
\langle \hat f_Q(\tau)\hat f_Q(\tau')\rangle_{\rm b}=\langle \hat f_P(\tau)\hat f_P(\tau')\rangle_{\rm b}=\lambda\kappa (2\bar n+1)\delta(\tau-\tau'),
\end{equation}
and $\langle [\hat f_Q(\tau),\hat f_P(\tau')]\rangle_{\rm b}=2i\lambda\kappa\delta(\tau-\tau')$. This commutation condition guarantees that the commutation relation $[Q,P]=i\lambda$ does not change in time. The noise correlators are understood here in the Stratonovich sense \shortcite{vanKampen_book}; in particular,  $\langle[\hat f_Q(\tau),P(\tau)]\rangle_{\rm b} = \langle[Q(\tau),\hat f_P(\tau)]\rangle_{\rm b}=i\lambda\kappa$.

Near a bifurcation point we can simplify eqn~(\ref{eq:q_Langevin}) using essentially the same approach as for classical systems \shortcite{Dykman1980}. We can change to operators $Q-Q_B$ and $P-P_B$; here $Q_B$ and $P_B$ are the classical values of $Q$ and $P$ at the bifurcation point given by the appropriate stationary solutions of the classical noise-free equations of motion (\ref{eq:classical_eom}). We can then further change from $Q-Q_B,P-P_B$ to $Q',P'$ by rotating coordinates in the $(Q,P)$ plane. We choose the angle of rotation in such a way that {\it at the bifurcation point} the equation for $\dot P'$ does not contain linear in $Q',P'$ terms. The equation of motion for $Q'$, on the other hand, has the form
\[\dot Q'=-A_{QQ}Q' +A_{QP}P'+({\rm nonlinear \; terms \; in}\, Q',P')+ \hat f_{Q'}.\]
Therefore, for small expectation values of $Q',P'$, the operator $P'$ is slowly varying in time compared to $Q'$. We note that, for an additively modulated oscillator, the rotation is not needed, $P'=P-P_B$, and for a parametrically modulated oscillator $Q_B=P_B=0$.

We will study the slow dynamics of  $P'$ for a small deviation $\eta$ of the control parameter from its bifurcation value. For $|\eta|\ll 1$, in dimensionless time $1/A_{QQ}\sim 1$ operator $Q'$ reaches its adiabatic form $\approx (A_{QP}/A_{PP})P' + {\mathcal O}(\eta)$, while $P'$ remains unchanged. The expression for $Q'$ can be then substituted into equation for $\dot P'$. The resulting equation for $\dot P'$ reads
\begin{equation}
\label{eq:slow_P}
\dot P'\approx B_0\eta +B_1\eta P' + B_2 P^{\prime \,2}+ B_3P^{\prime \,3} + \hat f_{P'}(\tau).
\end{equation}
Here, near the pitchfork bifurcation point for a parametrically modulated oscillator $B_0=B_2=0$ by symmetry and $B_{1,3}\sim 1$; near the saddle-node bifurcation point for an additively driven oscillator $B_{0,2}\sim 1$ while the terms $\propto \eta P',P^{\prime\,3}$ can be disregarded, cf. \shortciteN{Guckenheimer1987}. The explicit form of $B_{0,1,2,3}$ follows from eqns~(\ref{eq:q_Langevin}).

The correlator of $\hat f_P'$ is the same as of $\hat f_P$ in eqn~(\ref{eq:noise_correlators}), whereas $\hat f_{Q'}$ drops out from eqn~(\ref{eq:slow_P}), to the leading order in $\eta$ ($\hat f_{Q'}$ enters the equation for $\dot P'$ with a coefficient $\propto P'$). Therefore the $\delta$-correlated noise $\hat f_{P'}$ behaves as classical, as it commutes with itself, and then $P'$ behaves as a classical variable. However, the intensity of $\hat f_{P'}$ is $\propto 2\bar n +1$, so that the fluctuations still have quantum origin.

Equation (\ref{eq:slow_P}) can be written as $\dot P'=-\partial_{P'}U(P') + \hat f_{P'}(\tau)$. It maps the oscillator dynamics onto the dynamics of an overdamped classical Brownian particle in a potential $U$. The form of the potential depends on the nature of the bifurcation. For an additively modulated oscillator, where the stable and unstable states of forced vibrations merge for $\beta=\beta_B$, this potential in rescaled variables is shown in Fig.~\ref{fig:bif_potentials}(a), whereas for a parametrically modulated oscillator where the period-two states merge for $\mu=\mu_{B1}$, it is shown in Fig.~\ref{fig:bif_potentials}(b); the dynamics for small $\mu -\mu_{B2}$ is described by a potential of the opposite sign.

Switching from a metastable vibrational state corresponds to a quantum-activated escape from the corresponding minimum of the potential $U$. The rate of escape via quantum tunneling is exponentially smaller \cite{Dykman2007}. The activation exponent $R_A$ is determined by the height of the potential barrier, and therefore it displays a characteristic scaling dependence on the distance to the bifurcation point  $\eta=\beta-\beta_B$ or $\eta=\mu-\mu_B$. The results are summarized in Table~\ref{table:bifurcation}.
\noindent
\begin{table}[t]
\tableparts
{
\caption{Quantum-activated switching near bifurcation points, $W_{\rm sw}=\Omega_{\rm sw}\exp(-R_A/\lambda)$}
\label{table:bifurcation}
}
{\begin{tabular}{lll}
\hline
 & & \\[-8pt]
& Additive driving & Parametric driving\\ [3pt]
\hline
 & &\\[-5pt]
 Bifurcation points & $\beta_{B1,2}=\frac{2}{27}\left[1+9\kappa^{2}
\mp\left(1-3\kappa^{2}\right)^{3/2}\right]$ & $\mu_{B1,2}=\mp (1-\kappa^2)^{1/2}$\\[5pt]
Squared amplitude & &\\
at bifurcation points &
$(A^2_{\rm osc})_{B1,2}=\frac{1}{3}\left[2\pm(1-3\kappa^{2})^{1/2}\right]$ & $(A^2_{\rm osc})_{B1,2}=0$\\[5pt]
Distance to bifurcation & $\eta = \beta-\beta_B $ & $\eta = \mu - \mu_B$\\[8pt]
Activation energy $R_A$& $\frac{2\sqrt{2}}{3\kappa |b|^{1/2}\beta_B^{3/4}}\, |\eta|^{3/2}/(2\bar n + 1)$ & $\frac{1}{2}|\mu_B|\,\eta^2/(2\bar n + 1)$\\[8pt]
Prefactor $\Omega_{\rm sw}$ & $|\delta\omega|(b\eta/2)^{1/2}/\pi\beta_B^{1/4}$ & $\Gamma |\eta \mu_B|\frac{[1+\Theta(\mu_B)]}{2^{1/2}\pi\kappa^2}$\\[8pt]
Auxiliary parameter& $b=\beta_B^{1/2}[3(A^2_{\rm osc})_B-2]/2\kappa^2 $\\[8pt]
\hline
\end{tabular}
\vspace*{-4pt}
}
\end{table}
%

The scaling behavior of $\ln W_{\rm sw}$ with the distance to the bifurcation point\index{bifurcation!scaling} for classical oscillators has been seen for additive driving near the saddle-node  bifurcation points \shortcite{Siddiqi2006a,Stambaugh2006} and near the critical point ($\beta = 8/27, \kappa=1/\sqrt{3}$) where both stable vibrational states and the unstable state in Fig.~\ref{fig:phase_portrait}(a) merge together \shortcite{Aldridge2005}\footnote{The results on the switching rates for an additively modulated classical oscillator near the critical point \protect\shortcite{Dykman1980} extend to the quantum regime if one replaces $k_BT\to \hbar\omega_0(\bar n+1/2)$.}, and for parametrically modulated oscillators near the pitchfork bifurcations\index{bifurcation!pitchfork} (\shortciteNP{Chan2007}; see also the chapter by Chan and Stambaugh in this book). Recently the scaling behavior near the saddle-node bifurcation\index{bifurcation!saddle-node} and the characteristic temperature dependence $\ln W_{\rm sw}\propto (2\bar n + 1)^{-1}$  were found also in the quantum regime (\shortciteNP{Vijay2009}; see also the chapter by R.~Vijay {\it et al.}in this book). The results provide direct evidence in support of the mechanism of quantum activation.\index{quantum activation}

\section{Power spectra of modulated quantum oscillators}
\label{sec:q_heat_spectra}

Of significant interest are spectra of a resonantly modulated oscillator, including the power spectrum\index{power spectrum} and the spectrum of the response to an additional field \shortcite{Dykman1979a,Drummond1980c,Collett1985}. Among other characteristics, the power spectrum determines the emission spectrum of the oscillator and relaxation of a qubit coupled to the oscillator \shortcite{Serban2010}. Spectral measurements have been reported both in the classical \shortcite{Stambaugh2006a,Almog2007} and quantum regimes \shortcite{Wilson2010}; see also the chapters by Chan and Stambaugh and by Wilson {\it et al.}.

The spectra of interest are described by functions
\begin{eqnarray}
\label{eq:spectrum_general}
&\lal K,L\rar_{\omega}=\int\nolimits_0^{\infty}dte^{i\omega t}\lal K(t)L(0)\rar,&\nonumber\\
&\lal K(t)L(0)\rar = \frac{\omega_{\rm M}}{2\pi}\int\nolimits_0^{2\pi/\omega_{\rm M}}dt_i\langle [K(t+t_i)-\langle K(t+t_i)\rangle][L(t_i)-\langle L(t_i)\rangle]\rangle;&
\end{eqnarray}
recall that $\omega_{\rm M}$ is $\omega_F$ for additive and $\omega_F/2$ for parametric modulation; $\omega_{\rm M}$ is close to $\omega_0$. We will consider spectra near resonance, with $|\omega|\approx \omega_0$; the operators $K$ and $L$ will be the ladder operators $a$ or $a^{\dag}$. In particular, the peak in the spectrum of spontaneous radiation emission by the oscillator is determined by Re~$\lal a^{\dag},a\rar_{\omega}$ with $\omega\approx -\omega_0$, as in the absence of periodic modulation, cf. \shortciteN{Mandel1995}. A physical example is radiation from a nonlinear cavity, with the oscillator being the cavity mode modulated by an incident electromagnetic field \shortcite{Drummond1980c} or excited by modulating the boundary of the cavity \shortcite{Wilson2010}.

An additional weak resonant force $A'\exp(-i\omega t) +$~c.c. causes the oscillator to vibrate at frequencies $\omega$ and $2\omega_{\rm M}-\omega$. The vibrations at frequency $\omega$ are described by the scaled susceptibility $\chi(\omega)$,\index{susceptibility} which determines the corresponding displacement $\langle \delta q\rangle =(A'/2\omega_{\rm M})\chi(\omega)\exp(-i\omega t) +$~c.c.,
\begin{equation}
\label{eq:susceptibility_define}
\chi(\omega)= i\left[\lal a,a^{\dag}\rar_{\omega}-\lal a^{\dag},a\rar_{-\omega}^*\right].
\end{equation}
In the absence of modulation eqn~(\ref{eq:susceptibility_define}) goes over into the standard expression for the oscillator susceptibility. Function $\chi(\omega)$ includes an extra factor $2\omega_{\rm M}$ compared to the notation we used previously \shortcite{Dykman1994b}.

Spectra of a modulated oscillator have two major contributions. One comes from motion in the vicinity of stable vibrational states, where the oscillator spends much of the time, and the other comes from fluctuation-induced transitions between the states. We will consider them separately.

\subsection{Fluctuations about a stable vibrational state and local response}

The contribution to the oscillator spectra from motion near a stable vibrational state should be analyzed differently depending on the dimensionless oscillator relaxation rate $\kappa$. For $\kappa\gg \lambda$ (see below), one can linearize and then solve quantum equations of motion (\ref{eq:q_Langevin}) near the position $(Q_{\rm a},P_{\rm a})$ of a given stable state, similar to the classical case \shortcite{Dykman1979a,Dykman1994b} and to how it is done in the coherent state representation \shortcite{Drummond1980c}. Equivalently, one can linearize in $(Q-Q_{\rm a},P-P_{\rm a})$ the drift term in the master equation for the density matrix in the Wigner representation (\ref{eq:wigner}). The resulting contribution to the power spectrum is \shortcite{Serban2010,Dykman2011}
\begin{eqnarray}
\label{eq:spectra_linearization1}
{\rm Re}~\lal a,a^{\dag}\rar_{\omega}^{\rm (a)}\approx \frac{\kappa^2}{\Gamma}\frac{(\bar n +1)\left[\left(\nu -c_1\right)^2 + \kappa^2\right] + \bar n c_2}
{(\nu^2-\nu_{\rm a}^2)^2+4\kappa^2\nu^2}, \qquad \nu=\frac{\kappa}{\Gamma}(\omega-\omega_{\rm M}).
\end{eqnarray}
The parameters in eqn~(\ref{eq:spectra_linearization1}) are expressed in terms of the scaled squared vibration amplitude in the considered stable state $r_{\rm a}^2=Q_{\rm a}^2+P_{\rm a}^2$,
\begin{eqnarray}
\label{eq:spectral_parameters}
&&(\nu_{\rm a}^2)_{\rm add}=\kappa^2+3r_{\rm a}^4-4r_{\rm a}^2+1,\qquad (c_1)_{\rm add}=1-2r_{\rm a}^2,
\qquad (c_2)_{\rm add}=r_{\rm a}^4,\nonumber\\
&&(\nu_{\rm a}^2)_{\rm par}=4r_{\rm a}^2(r_{\rm a}^2-\mu),\qquad (c_1)_{\rm par}=\mu-2r_{\rm a}^2,
\qquad (c_2)_{\rm par}=\kappa^2+\mu^2,
\end{eqnarray}
where subscripts ``add" and ``par" refer to additive and parametric modulation,\index{modulation!additive}\index{modulation!parametric} respectively; in eqn~(\ref{eq:spectra_linearization1}) we use the factor $\Gamma/\kappa$ as the frequency scale, this factor is independent of the decay rate $\Gamma$. In the limit of small damping, $\kappa\to 0$, $\nu_{\rm a}$ goes over into the dimensionless frequency $\nu_0$ of vibrations about the attractor, eqn~(\ref{eq:squeezed_operators}).

The contribution of small-amplitude fluctuations to Re~$\lal a^{\dag},a\rar_{-\omega}$ for $\omega$ close to $\omega_0$ is also given by eqn~(\ref{eq:spectra_linearization1}) in which one should interchange $\bar n\leftrightharpoons\bar n+1$. This relation, together with eqn~(\ref{eq:susceptibility_define}), give the imaginary part of the susceptibility $\chi(\omega)$. In evaluating these local contributions we replace in the definition of the power spectrum (\ref{eq:spectrum_general}) $\langle a(t)\rangle \to \langle a(t)\rangle^{\rm (a)}=(2\lambda)^{-1/2}(Q_{\rm a}+iP_{\rm a})\exp(-i\omega_{\rm M}t)$.

The susceptibility for not too small $\kappa$ has the same form in the quantum and classical case. It can be obtained by adding to the linearized in $Q-Q_{\rm a}, P-P_{\rm a}$ eqns~(\ref{eq:q_Langevin}) a weak-driving term $\propto A'$, which oscillates in the rotating frame at frequency $\omega-\omega_{\rm M}$. One then finds the solution that oscillates at this frequency. Since the equations are linear, the noise term, which is the only term that has a different form in the classical and quantum case, drops out on averaging, giving for $\kappa\gg \lambda$
\begin{equation}
\label{eq:suscept_attractor1}
\chi_{\rm a}(\omega) \approx i\frac{\kappa}{\Gamma}\frac{\kappa-i(\nu - c_1)}
{\nu_{\rm a}^2-\nu^2-2i\kappa\nu},\qquad \nu=\frac{\kappa}{\Gamma}(\omega-\omega_{\rm M}).
\end{equation}

For underdamped vibrations about the stable state in the rotating frame, where $\kappa\ll \nu_0$ (but still $\kappa\gg \lambda$), the spectra (\ref{eq:spectra_linearization1}) and Im~$\chi_{\rm a}(\omega)$ have peaks at frequency detuning $\omega-\omega_{\rm M} =\pm (\Gamma/\kappa)\nu_0$. There are generally two peaks on the opposite sides of $\omega_{\rm M}$ and they generally have different amplitudes. Their shape is close to Lorentzian, with halfwidth $\Gamma$. As the ratio $\nu_0/\kappa$ decrease the peaks start overlapping.

\begin{figure}[h]
\begin{center}
\includegraphics[width=4.0 cm]{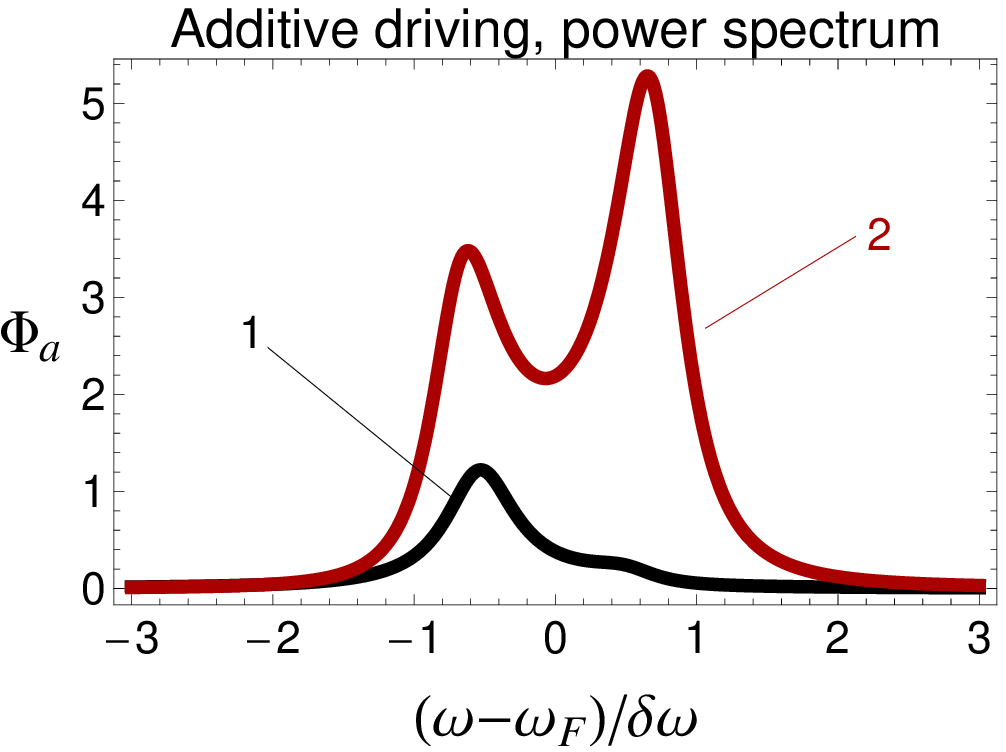} \hfill
\includegraphics[width=4.2 cm]{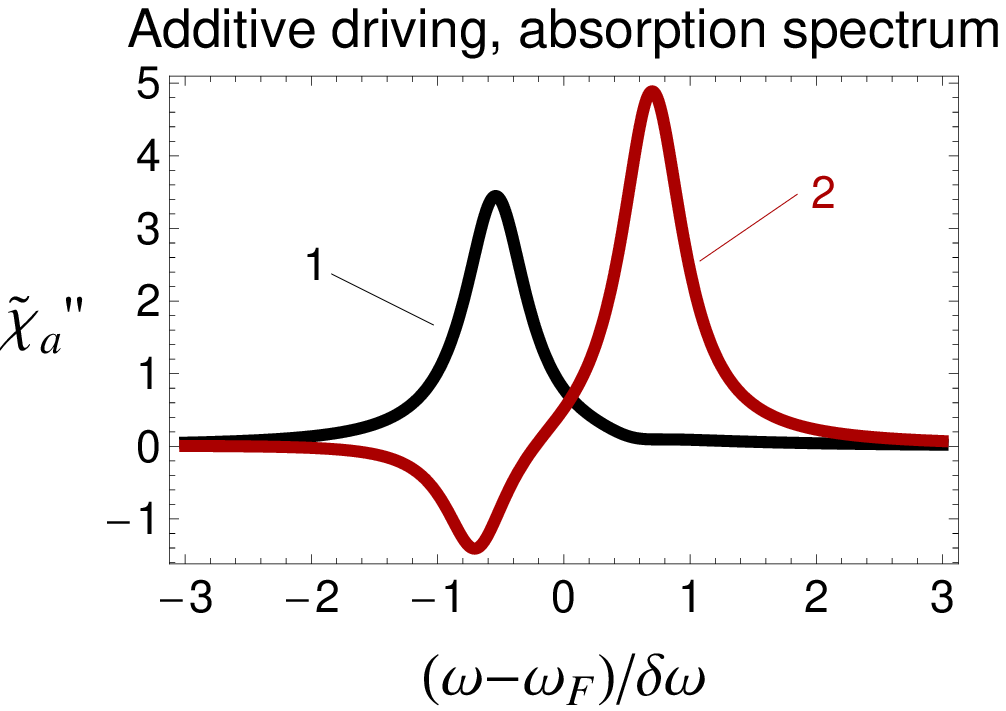}\hfill
\includegraphics[width=4.0 cm]{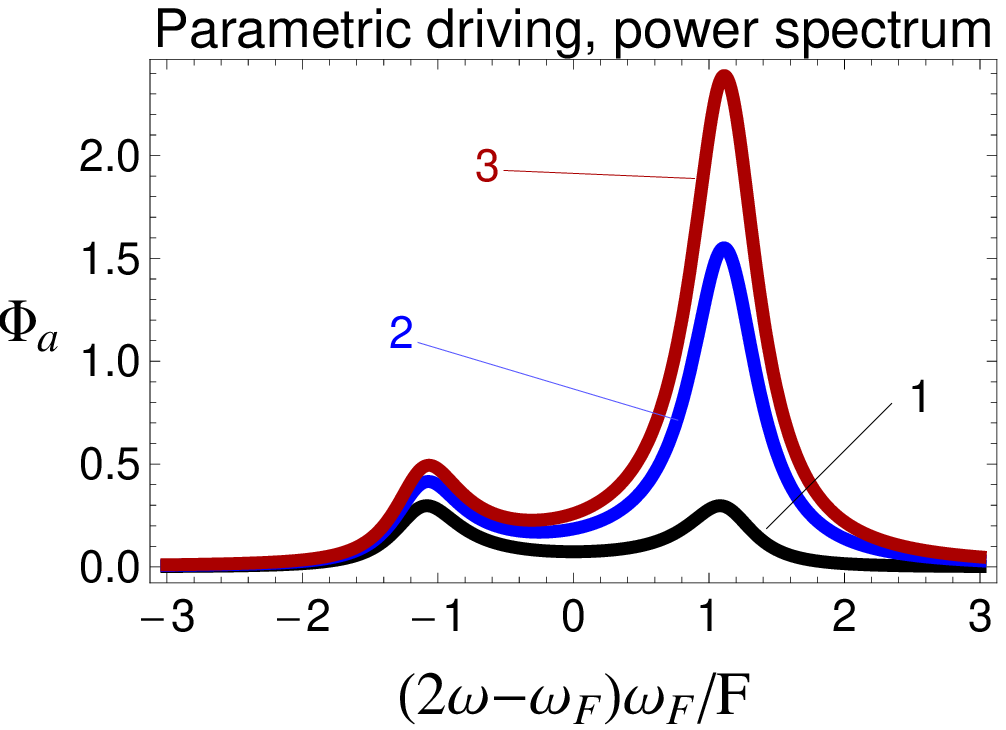}
\end{center}
\caption{The contributions to the scaled power spectra $\Phi_{\rm a}(\omega)={\rm Re}~(\kappa/\Gamma)\lal a^{\dag},a\rar_{-\omega}^{\rm (a)}$ and the susceptibility, $\tilde\chi''_{\rm a}(\omega)={\rm Im}~(\kappa/\Gamma)\chi_{\rm a}(\omega)$ from fluctuations about stable vibrational states for $\kappa \gg \lambda$. The scaled decay rate is $\kappa=0.3$. For the additively driven oscillator, curves 1 and 2 refer to the small- and large-amplitude vibrational states; the parameters are $\beta=0.15$ and $\bar n=0.3$. Curves 1 to 3 for the parametrically driven oscillator refer to $\mu=-0.6$ and $\bar n=0,0.3,0.5$.}
\label{fig:linearized_spectra}
\end{figure}

The spectra are illustrated in Fig.~\ref{fig:linearized_spectra}. As seen from the figure and eqn~(\ref{eq:spectra_linearization1}), the emission spectrum $\lal a^{\dag},a\rar_{-\omega}^{\rm (a)}$ is symmetric for $\bar n=0$; this holds for both additive \shortcite{Drummond1980c} and parametrically modulated oscillators. The onset of emission for $\bar n=0$ is related to quantum heating. In terms of quantum optics, if the modulating field is electromagnetic radiation and one considers emission of photons by the modulated oscillator, the emission can be thought of as resulting from multi-wave parametric process. Both the emission intensity and the positions of the spectral peaks depend on the modulation strength in a complicated way.

An interesting feature of the susceptibility seen from Fig.~\ref{fig:linearized_spectra} is that Im~$\chi(\omega)$ can become negative \shortcite{Dykman1979a}. In the corresponding frequency range an additional weak field is amplified by the strong field, which can be also considered as a parametric multi-wave process. The amplification occurs in spite the fact that the absorption coefficient integrated over the whole spectrum is positive: from eqn~(\ref{eq:suscept_attractor1}) $\int d\omega \chi_{\rm a}''(\omega)=\pi$. This sum rule holds because, for any Fock state $|N\rangle$ (cf. Fig.~\ref{fig:relaxation_sketch}), an induced dipolar transition up in energy has a larger amplitude [$\propto (N+1)^{1/2}$] than down in energy ($\propto N^{1/2}$). For a modulated nonlinear oscillator, the spectral regions of absorption and amplification are separated; weak field amplification \index{weak field amplification} generally occurs for both additive and parametric modulation.

\subsection{Fine structure of the spectra and the interference of transitions}
\label{subsec:fine_structure}

\begin{figure}[h]
\begin{center}
\includegraphics[width=3.1 cm]{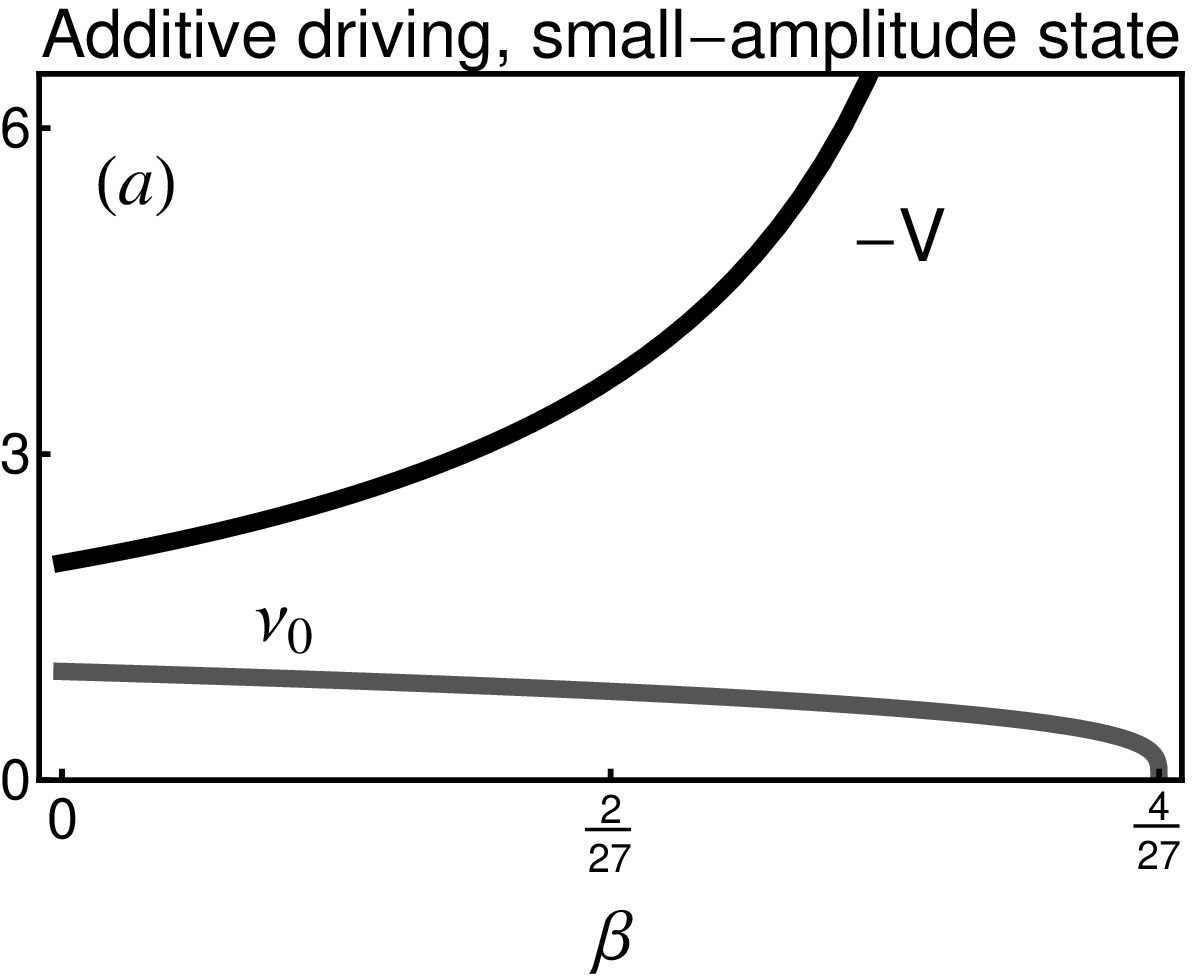}\hfill
\includegraphics[width=3.2 cm]{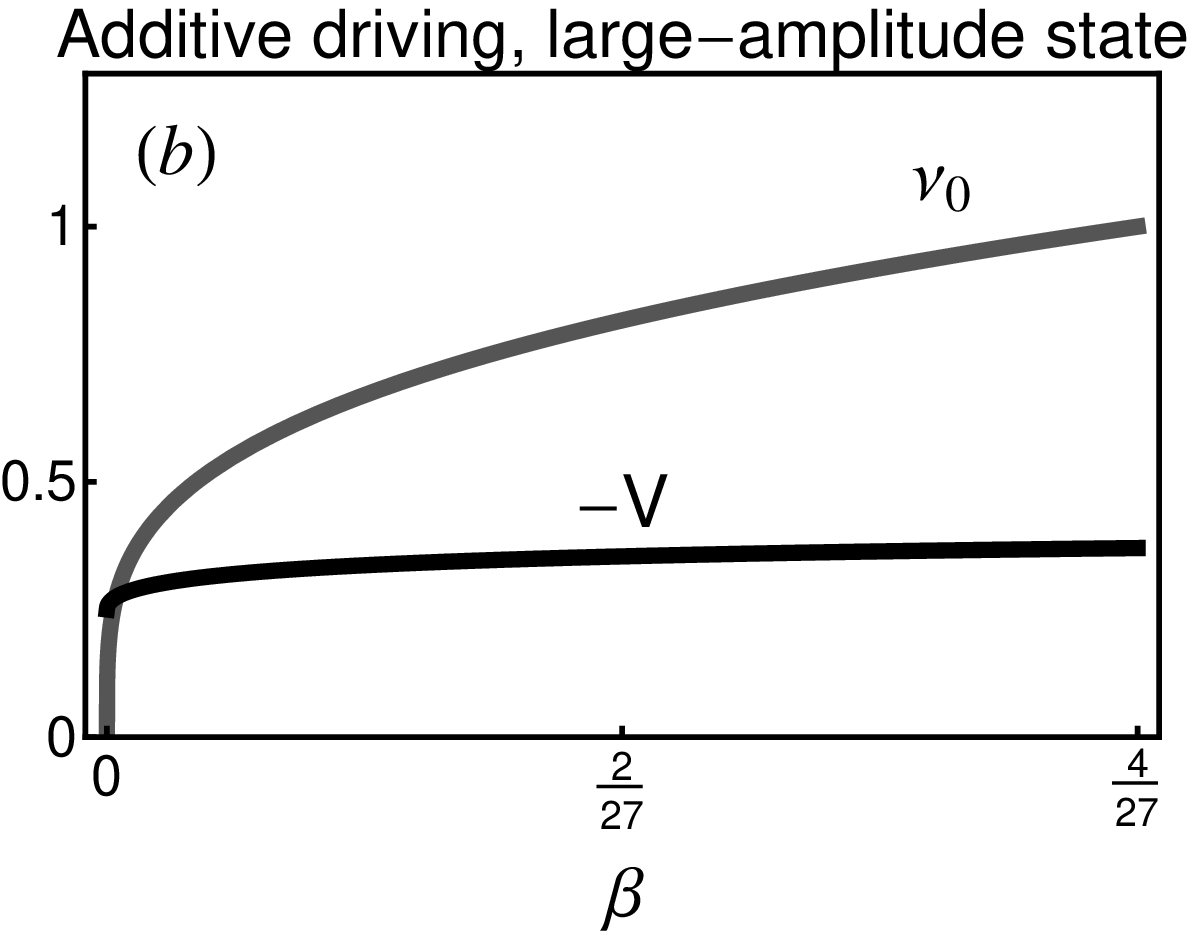}\hfill
\includegraphics[width=3.2 cm]{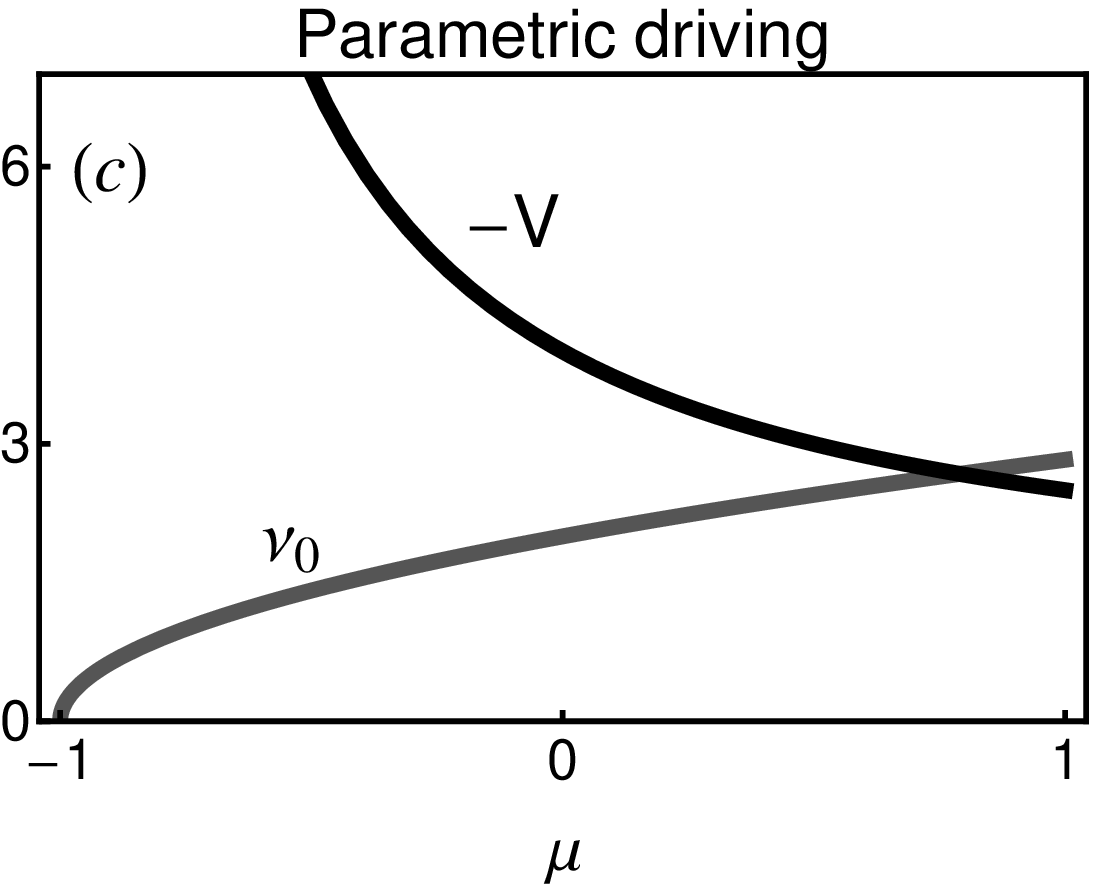}
\end{center}
\caption{The scaled eigenfrequency of vibrations about the stable states in the rotating frame $\nu_0$ and the parameter of nonequidistance of quasienergy levels $V$, eqn~(\protect\ref{eq:nonlinear_g_n}), for additive and parametric modulation.}
\label{fig:nonequidist_parameters}
\end{figure}

\begin{figure}[t]
\begin{center}
\includegraphics[width=2.1 cm]{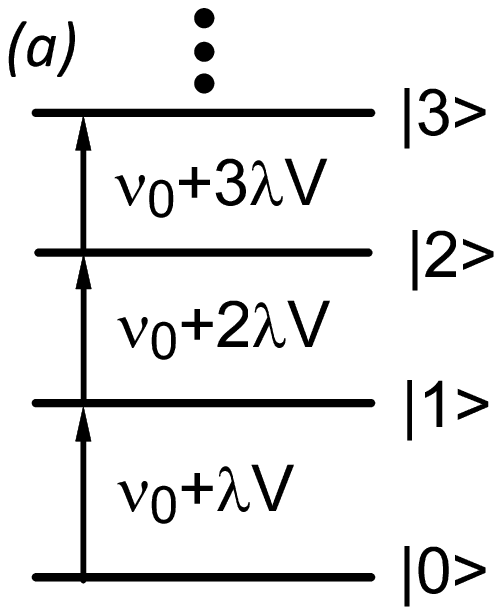}\hspace{2cm}
\includegraphics[width=4.0 cm]{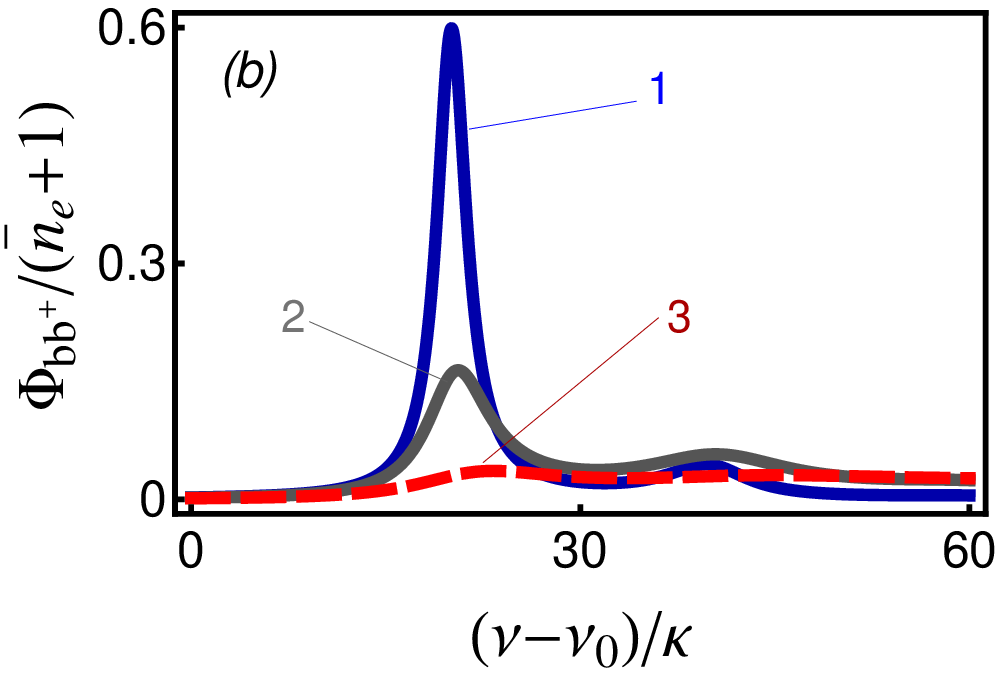}
\end{center}
\caption{(a) A sketch of the quasienergy spectrum and dimensionless transition frequencies of weakly nonlinear vibrations about the stable state in the rotating frame for $g_{QQ}>0$. (b) The power spectrum of the auxiliary oscillator near resonant frequency $\nu_0$ as given by eqn~(\protect\ref{eq:partial_general}) for a comparatively large ratio of the level nonequidistance to the decay rate, $\vartheta = 10$ and for $g_{QQ}>0$. The curves 1 to 3 refer to $\bar n_e= 0.1,0.5$ and 1.5. }
\label{fig:fine_structure}
\end{figure}

\index{power spectrum!fine structure}Interesting quantum effects emerge in the spectra for small $\kappa$, where not only $\kappa\ll \nu_0$, but also $\kappa\lesssim \lambda\ll \nu_0$. In the whole range $\kappa\ll \nu_0$ the stable states of the oscillator are at the extrema of $g(Q,P)$. However, the approach of the previous section does not apply for $\kappa\lesssim \lambda$. To consider the range $\kappa\lesssim \lambda$, one has to take into account the nonequidistance of the quasienergy levels $g_n$, which results from the nonlinearity of vibrations about the extrema. For a given extremum at $Q=Q_0,P=0$, to the leading order in the nonlinearity
\begin{equation}
\label{eq:nonlinear_g_n}
g_n\approx g_0+\left[\lambda\nu_0(n+1/2) + \lambda^2Vn(n+1)/2\right]{\rm sgn}g_{QQ},
\end{equation}
where $g_0=g(Q_0,0)$. Parameter $V=\nu_0(d\nu/dg)_{g_0}$ is determined by the slope of the dimensionless frequency $\nu(g)$ of vibrations in the rotating frame with given $g$; the quantum correction $\sim \lambda$ to the frequency $\nu_0=\nu(g_0)$ is assumed to be incorporated. The parameters $\nu_0$ and $V$ are plotted in Fig.~\ref{fig:nonequidist_parameters}. For small $\lambda$ the vibration nonlinearity is small, $\lambda |V|\ll \nu_0$. The quasienergy spectrum (\ref{eq:nonlinear_g_n})\index{quasienergy!spectrum} is sketched in Fig.~\ref{fig:fine_structure}(a).

The power spectrum of the oscillator in a given stable state is formed by transitions $|n\rangle \to |n\pm 1\rangle$ between neighboring quasienergy states in Fig.~\ref{fig:fine_structure}(a). One might expect that the spectrum is then a superposition of partial spectra that correspond to individual transitions, with width determined by the reciprocal lifetime of the respective quasienergy states. However,  to spectrally resolve the transitions one has to wait for time $\gtrsim (\kappa/\Gamma)|\lambda V|^{-1}$. If this time becomes comparable or smaller than the lifetime, the transitions are not independent, the transition amplitudes interfere. The overall oscillator spectrum is then formed by many interfering transitions.

The typical number of states that contribute to the spectrum is determined by the effective Planck number $\bar n_e$, eqn~(\ref{eq:n_e}). We assume that $\lambda |V|\bar n_e\ll \nu_0$, so that all dimensionless transition frequencies are close to $\pm \nu_0$. Therefore the spectrum has well-separated peaks at $\omega-\omega_{\rm M}\approx \pm (\Gamma/\kappa)\nu_0$. Near the peaks, eqn~(\ref{eq:spectrum_general}) can be simplified using the interrelation (\ref{eq:squeezed_operators}) between operators $Q,P$ and operators $b,b^{\dag}$ that describe vibrations about the stable state in the rotating frame. The operators $b,b^{\dag}$ are the ladder operators of an auxiliary oscillator in thermal equilibrium, with scaled eigenfrequency $\nu_0$, energy spectrum (\ref{eq:nonlinear_g_n}), and temperature ${\mathcal T}_e$. The power spectrum of this oscillator is
\begin{equation}
\label{eq:b_correlators}
\Phi_{bb^{\dag}}(\nu)={\rm Re}\int\nolimits_0^{\infty} dt e^{i\nu\tau}\langle b(\tau)b^{\dag}(0)\rangle  = e^{\lambda\nu_0/{\mathcal T}_e}\,{\rm Re} \int\nolimits_0^{\infty} dt e^{-i\nu\tau}\langle b^{\dag}(\tau)b(0)\rangle.
\end{equation}
Equation (\ref{eq:b_correlators}) is written for $|\nu-\nu_0{\rm sgn}g_{QQ}|\ll \nu_0$ where $\Phi_{bb^{\dag}}(\nu)$ has a narrow peak. The occurrence of such a  peak can be understood by noticing that,  if one disregards nonlinearity and decay, $b(\tau)=\exp(-i\nu_0\tau\,{\rm sgn}g_{QQ})b(0)$, and then $\Phi_{bb^{\dag}}(\nu)$ becomes a $\delta$-function. The nonlinearity and decay lead to broadening of the $\delta$-function.

From eqns~ (\ref{eq:squeezed_operators}) and (\ref{eq:spectrum_general}), near the peak of the power spectrum of the original modulated oscillator on the high-frequency (low-frequency, for ${\rm sgn}g_{QQ}<0$) side of the forced-vibration frequency, $\omega\approx \omega_{\rm M}+(\Gamma/\kappa)\nu_0\, {\rm sgn}g_{QQ}$,
\begin{equation}
\label{eq:spectra_interrelation_highf}
\frac{\kappa}{\Gamma}{\rm Re}~\lal a,a^{\dag}\rar_{\omega}^{\rm (a)}\approx \frac{\kappa}{\Gamma}e^{\lambda\nu_0/{\mathcal T}_e}{\rm Re}~\lal a^{\dag},a\rar_{-\omega}^{\rm (a)} \approx \cosh^2\varphi_*\Phi_{bb^{\dag}}(\nu),
\end{equation}
where, as in eqns~(\ref{eq:spectra_linearization1}) and (\ref{eq:suscept_attractor1}), $\nu=(\kappa/\Gamma)(\omega-\omega_{\rm M})$.

Similarly, near the peak in the power spectrum on the low-frequency (high-frequency, for ${\rm sgn}g_{QQ}<0$) side of $\omega_{\rm M}$, where  $\nu=(\kappa/\Gamma)(\omega-\omega_{\rm M})$ is close to $-\nu_0\,{\rm sgn}g_{QQ}$,
\begin{equation}
\label{eq:spectra_interrelation_lowf}
\frac{\kappa}{\Gamma}e^{\lambda\nu_0/{\mathcal T}_e}{\rm Re}~\lal a,a^{\dag}\rar_{\omega}^{\rm (a)}\approx \frac{\kappa}{\Gamma}{\rm Re}~\lal a^{\dag},a\rar_{-\omega}^{\rm (a)} \approx \sinh^2\varphi_*\Phi_{bb^{\dag}}(-\nu).
\end{equation}

Equations (\ref{eq:susceptibility_define}), (\ref{eq:spectra_interrelation_highf} ), and (\ref{eq:spectra_interrelation_lowf}) also describe resonant peaks in the oscillator absorption spectrum Im~$\chi_{\rm a}(\omega)$. The shapes of the peaks are determined by function $\Phi_{bb^{\dag}}(\nu)$. The peak near frequency $\omega_{\rm M} + (\Gamma/\kappa)\nu_0\,{\rm sgn}g_{QQ}$ corresponds to absorption of the additional field, Im~$\chi_{\rm a}(\omega) >0$, whereas the one near $\omega_{\rm M} - (\Gamma/\kappa)\nu_0\,{\rm sgn}g_{QQ}$, with Im~$\chi_{\rm a}(\omega) <0$,  corresponds to amplification and has a smaller area.

Equations (\ref{eq:spectra_interrelation_highf}) and (\ref{eq:spectra_interrelation_lowf}) reduce the problem of the spectra of modulated oscillator for weak damping, $\kappa\ll \nu_0$, to calculating the power spectrum of an auxiliary equilibrium oscillator $\Phi_{bb^{\dag}}(\nu)$ \shortcite{Dykman2011}. This problem was discussed previously \cite{DK_review84}. The result for the spectrum near its maximum can be presented in the form
\begin{eqnarray}
\label{eq:partial_general}
&\Phi_{bb^{\dag}}(\nu) = (\bar n_e+1){\rm Re}~\sum\nolimits_{n=1}^{\infty}\phi{}(n,\nu);&\\
&\phi{}(n,\nu)=
 4n(\Lambda-1)^{n-1}(\Lambda+1)^{-(n+1)}\left[\kappa (2 \aleph n -1)-i(\nu-\nu_0\,{\rm sgn}g_{QQ})\right]^{-1},&\nonumber
\end{eqnarray}
where $\Lambda= \aleph^{-1}\left[1 +i\vartheta(2\bar{n}_e+1)\right]$ and
$\aleph  = \left[1 + 2i \vartheta(2\bar{n}_e+1)-\vartheta^2\right]^{1/2} \;({\rm Re}~\aleph > 0)$. The parameter $\vartheta =(\lambda V/2\kappa)\,{\rm sgn}g_{QQ}$ is determined by the interrelation between the nonequidistance of the transition frequencies in Fig.~\ref{fig:fine_structure}(a) and the decay broadening $\kappa $ of the quasienergy levels,.

Equation (\ref{eq:partial_general}) represents the spectrum as a sum of  effective partial spectra Re~$\phi{}(n,\nu)$ that can be provisionally associated with transitions $|n-1\rangle\to |n\rangle$ between the quasienergy levels in Fig.~\ref{fig:fine_structure}(a). Functions $\phi{}(n,\nu)$ depend on two parameters, $\vartheta$ and $\bar n_e$.
The form of $\phi{}(n,\nu)$ is particularly simple for a comparatively large level nonequidistance or small damping, $\lambda |V|\gg \kappa $, in which case
\begin{eqnarray}
\label{eq:fine_structure}
\phi{}(n,\nu)\approx  \frac{n}{(\bar n_e+1)^2}\frac{\exp[-\lambda\nu_0(n-1)/{\mathcal T}_e]}{\kappa_n  -i\left[\nu-\nu(g_{n-1})\right]},\quad |\vartheta|\equiv \lambda \frac{|V|}{2\kappa}\gg 1.
\end{eqnarray}
Here, $\nu(g_{n-1})=(\nu_0+\lambda Vn){\rm sgn}g_{QQ}$ is the frequency of the $|n-1\rangle \to |n\rangle$ transition in Fig.~\ref{fig:fine_structure}(a) and $\kappa_n   = \kappa [2n(2\bar n_e+1)-1]$ is the total halfwidth of levels $|n-1\rangle$ and $|n\rangle$. Therefore Re~$\phi{}(n,\nu)$ has a conventional form of a partial spectrum.

For $|\vartheta|\gg 1$ the overall spectrum $\Phi_{bb^{\dag}}(\nu)$ has a fine structure. The intensities of the individual lines (\ref{eq:fine_structure}) immediately give the effective quantum temperature ${\mathcal T}_e$.\index{effective temperature} However, the fine structure is pronounced only in a limited range of the effective Planck numbers $\bar n_e$. This is seen from Eq.~(\ref{eq:fine_structure}). For $\bar n_e\ll 1$ only $\phi{}(1,\nu)$ has an appreciable intensity while Re~$\kappa\phi{}(n,\nu)\ll 1$ for $n>1$. On the other hand, for large $\bar n_e$ the linewidth $\kappa_n$ becomes large and spectral lines with different $n$ overlap, starting with large $n$. Not only are they overlapping, but their shape is also changed compared to eqn~(\ref{eq:fine_structure}) due to the interference of transitions.\index{interference of transitions} The evolution of the fine structure with varying $\bar n_e$ as given by eqn~(\ref{eq:partial_general}) is illustrated in Fig~\ref{fig:fine_structure}(b).

As $|\vartheta|$ decreases all partial spectra start to overlap, and for $|\vartheta|\lesssim 1$ they can no longer be identified. As a result of the interference of transitions, in the limit $|\vartheta|\to 0$ we have $\phi{}(n,\nu)\propto \delta_{n,1}$, and the spectrum has the form of a single Lorentzian peak of dimensionless halfwidth $\kappa$,
$\Phi_{bb^{\dag}}(\nu)=(\bar n_e+1)\kappa\left[\kappa^2 +(\nu-\nu_0{\rm sgn}g_{QQ})^2\right]^{-1}$. This expression, with account taken of eqn~(\ref{eq:spectra_interrelation_highf}), agrees with eqn~(\ref{eq:spectra_linearization1}) in the range $\lambda |V|\ll \kappa\ll \nu_0$ where both apply.
Generally, because of the nonlinearity of the auxiliary oscillator, the shape of the spectrum depends on $\bar n_e$ even where there is no fine structure. The spectrum is non-Lorentzian and displays a characteristic asymmetry for large $\bar n_e$, where $|\vartheta|\bar n_e > 1$. This asymmetry is described by eqn~(\ref{eq:partial_general}) and provides a way of determining quantum temperature.

Interference of transitions occurs also in various quantum oscillators and oscillator-type systems in the absence of modulation, from localized vibrations in solids to large spins in strong magnetic fields and Josephson-junction based systems, to mention but a few. In all these systems the spectra strongly depend on the interrelation between the level nonequidistance and the decay rate. Interference of transitions is important also for classical vibrational systems with fluctuating frequency, like nano- and micromechanical resonators with a fluctuating number and/or positions of attached molecules \shortcite{Vig1999,Yang2011}. The spectra of such systems can also be asymmetric and display a fine structure \shortcite{Dykman2010}. On the formal side, these recent results indicate that, for different systems and physical mechanisms, interference of transitions can be described by linear equations for coupled partial spectra. These equations are convenient for numerical solution.

\subsection{Supernarrow spectral peaks}
\label{subsec:supernarrow}

\index{supernarrow spectral peak}Along with fluctuations about stable vibrational states, quantum noise leads to occasional interstate switching discussed in Sec.~\ref{sec:quantum_activation}. Important manifestations of switching are additional peaks in the oscillator power spectrum and susceptibility. The peaks are centered at the forced-vibration frequency $\omega_{\rm M}$ and are supernarrow in the sense that their width is much smaller than the oscillator decay rate $\Gamma$.

To describe the peak in the power spectrum we note that the populations $w_{1}$ and $w_2=1-w_1$ of the stable vibrational states 1 and 2 satisfy the balance equation
\begin{equation}
\label{eq:occupation_balance_eq}
dw_{1}/dt = - W(w_1 - \bar w_1),\qquad W = W_{\rm sw}^{(12)}+W_{\rm sw}^{(21)}, \qquad \bar w_1= W_{\rm sw}^{(21)}/W,
\end{equation}
where $W_{\rm sw}^{(ij)}$ is the rate of switching from $i$th to $j$th state and $\bar w_1$ is the mean state population. Equation (\ref{eq:occupation_balance_eq}) has the same form as in the classical case, except that the switching rates are determined by quantum fluctuations.

Fluctuations of the state populations lead to fluctuations of the expectation values of the operators $a(t), a^{\dagger}(t)$ averaged over time $\sim \Gamma^{-1}$, which switch between their stable-states values $a_i(t), a_i^{\dagger}(t)$, $i=1,2$; in the lab frame in the $i$th stable state $a_i(t)= a_i\exp(-i\omega_{\rm M}t)$ with $a_i=(2\lambda)^{-1/2}(Q_{{\rm a}_i} + iP_{{\rm a}_i})$ \shortcite{Dykman2011}. From eqn~(\ref{eq:occupation_balance_eq}), similar to the case of a classical oscillator \shortcite{Dykman1994b},  for the contribution from interstate switching to the power spectrum we obtain
\begin{equation}
\label{eq:power_switching}
{\rm Re}\lal a,a^{\dagger}\rar_{\omega}^{\rm (sw)}\approx {\rm Re}\lal a^{\dagger},a\rar_{\omega}^{\rm (sw)} \approx |a_1 - a_2|^2 \bar w_1 \bar w_2 W/[W^2 + (\omega-\omega_{\rm M})^2].
\end{equation}

The typical width of the spectral peak (\ref{eq:power_switching}) is given by the total switching rate $W$, it is exponentially smaller than the decay rate, cf. eqn~(\ref{eq:define_R_A}). The intensity (area) of the peak is determined by the factor $\bar w_1 \bar w_2$. For a parametrically modulated oscillator for $\mu<\mu_{B2}$ the populations of the states are equal by symmetry and this factor is equal 1/4. A peak in the power spectrum related to interstate transitions was seen in the radiation from a parametrically modulated microwave cavity \shortcite{Wilson2010}.

For additive driving, on the other hand, the populations $\bar w_{1}$ and $\bar w_{2}$ are exponentially different, and thus $\bar w_1\bar w_2$ is exponentially small everywhere except for a narrow parameter range where the switching rates $W_{\rm sw}^{(12)}$ and $W_{\rm sw}^{(21)}$ are almost equal. This range corresponds to a smeared kinetic ``phase transition" \shortcite{Bonifacio1978,Dykman1979a,Lugiato1984}, with the stable states playing the role of coexisting phases in a thermodynamic system. The onset of the supernarrow peak (\ref{eq:power_switching}) is an indicator of the transition \shortcite{Dykman1994b}. For a classical oscillator, such peak was observed by \shortciteN{Stambaugh2006a}.

The susceptibility of the oscillator also displays a supernarrow peak. The analysis of this peak for additive driving is similar to that for a classical oscillator \shortcite{Dykman1979a,Dykman1994b}. One notices that an extra force $A'\exp(-i\omega t)+$~c.c. with frequency very close to the strong-force frequency, $|\omega-\omega_F|\ll \Gamma$, can be described by making the parameter $\beta$ in eqn~(\ref{eq:g_additive}) slowly dependent on time, $\beta \to \beta(t)= \beta+(2\beta/A)\{A'\exp[-i(\omega-\omega_F)t] + {\rm c.c.}\}+\ldots$ [we do not consider terms at the mirror frequency $2\omega_F-\omega$ in $\beta(t)$]. One can then think of the switching rates becoming parametrically dependent on time via $\beta(t)$. From eqn~(\ref{eq:define_R_A}), the major part of this time dependence comes from the modulation of the activation energies $R_{A1}$ and $R_{A2}$ for switching from states 1 and 2, respectively. The state populations also become time dependent, which gives the switching-induced contribution to the susceptibility
\begin{equation}
\label{eq:switch_suscept}
\chi^{\rm (sw)}(\omega)\approx \bar w_1 \bar w_2 \frac{W}{W-i(\omega-\omega_F)}\frac{a_1-a_2}{\omega_F-\omega_0} (2\beta/\lambda)^{1/2}\left(\partial_{\beta}R_{A1}-\partial_{2\beta}R_{A2}\right).
\end{equation}

Function $\chi^{\rm (sw)}(\omega)$ displays resonant structure in a frequency range determined by the switching rate, $|\omega-\omega_F|\lesssim W$. The amplitude of $\chi^{\rm (sw)}(\omega)$ is proportional to a large factor $\sim R_{A1,2}/\lambda \gg 1$. It also contains factor $\bar w_1 \bar w_2$, which steeply depends on the distance to the kinetic phase transition. For a classical oscillator, the switching-induced peak of the response has been seen in the experiment \shortcite{Chan2006,Almog2007}. The susceptibility of a parametrically modulated oscillator also displays a supernarrow peak for $\omega =\omega_{\rm M}$ with amplitude $\propto R_A/\lambda\gg 1$. It was considered previously for a classical oscillator \shortcite{Ryvkine2006a}.

The amplitude of the supernarrow peak of the susceptibility increases with decreasing temperature, as the ratio $R_A/\lambda$ increases. The very onset of this peak for low temperatures, as well as the supernarrow peaks in the power spectrum, is due to quantum fluctuation-induced interstate switching.

\section{Nonresonant modulation: oscillator heating and cooling}
\label{sec:cooling}

\index{modulation!nonresonant}\index{heating, modulation induced}\index{cooling, modulation induced}We now briefly discuss oscillator dynamics in the presence of a moderately strong nonresonant modulation, where the modulation frequency $\omega_F$ is significantly different from the oscillator eigenfrequency $\omega_0$. The dynamics of a modulated linear oscillator with linear in coordinate $q$ and momentum $p$ interaction with a thermal bath (\ref{eq:linear_relaxation}) was studied earlier \shortcite{Schwinger1961,Zeldovich1970}. It can change dramatically if the interaction is nonlinear  \shortcite{Dykman1978} and/or the bath is also modulated.

A major effect of nonresonant modulation is the creation of a new relaxation mechanism.  The relaxation discussed earlier in this chapter, see Fig.~\ref{fig:relaxation_sketch}, was due to oscillator transitions between its neighboring energy levels. In a transition, the energy $\approx \hbar\omega_0$ is transferred to or taken from excitations in the bath. In the absence of modulation this imposes on the oscillator the thermal distribution of bath excitations with energy $\hbar\omega_0$, leading to the Boltzmann distribution $\rho_{NN}\propto \exp(-N\hbar\omega_0/k_BT)$ over Fock states $|N\rangle$.

\begin{figure}[h]
\begin{center}
\includegraphics[width=4.5 cm]{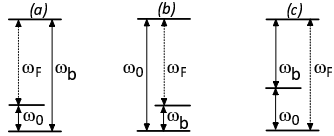}
\end{center}
\caption{A sketch of modulation-induced relaxation processes leading to cooling (a), heating (b), and population inversion (c) in the oscillator; $\omega_0$, $\omega_F$, and $\omega_{\rm b}$ are the oscillator frequency, the modulation frequency, and the frequency of the bath excitation, respectively. A sufficiently strong modulation imposes on the oscillator the probability distribution of the bath modes involved in the scattering in (a) and (b) and leads to population inversion of low-lying energy levels and self-sustained vibrations of the oscillator in (c).}
\label{fig:heating_cooling}
\end{figure}

In the presence of nonresonant modulation, in a bath-induced interlevel transition the part $\hbar\omega_F$ of the energy can come from the modulation, see Fig.~\ref{fig:heating_cooling}. The energy of the involved bath excitations is then $\hbar|\omega_0\pm\omega_F|$. These excitations should impose on the oscillator their thermal distribution. It corresponds to oscillator temperature $T^*=T\omega_0/(\omega_0\pm\omega_F)$. This temperature is positive if transitions $|N\rangle\to |N-1\rangle$ correspond to emission of bath excitations and are thus more probable than transitions $|N-1\rangle\to |N\rangle$, see Fig.~\ref{fig:heating_cooling}(a), (b). If, on the other hand, bath excitations are emitted in transitions $|N-1\rangle\to |N\rangle$, then $T^*$ becomes negative. This case is sketched in Fig.~\ref{fig:heating_cooling}(c). It corresponds to relaxation processes in which the energy of a ``photon" of the modulation goes into excitation of the oscillator and the bath.

Interestingly, the distribution of a nonresonantly modulated quantum oscillator over Fock states can remain of the Boltzmann form in a broad parameter range \shortcite{Dykman1978}. We will show this first for the case where the Hamiltonian of the interaction with the bath along with the linear in $a,a^{\dag}$ term (\ref{eq:linear_relaxation}) has a quadratic term\footnote{The interaction (\ref{eq:quadratic_coupling}) leads to nonlinear friction \shortcite{DK_review84},\index{nonlinear friction} which can play an important role in oscillator dynamics, see the chapter by Moser {\it et al.}, and in particular determine the characteristics of self-sustained vibrations; we do not discuss this dissipation mechanism here.}
\begin{equation}
\label{eq:quadratic_coupling}
H_i^{(2)}=q^2h_{\rm b}^{(2)}.
\end{equation}
The modulation term in the Hamiltonian of the isolated oscillator (\ref{eq:H_0(t)})  is
\begin{equation}
\label{eq:nonresonant_H_F}
H_F = H_{\rm add}=-qA{}\cos\omega_Ft;\qquad\omega_F, |\omega_F-\omega_0|\gg \Gamma,|\gamma|\langle q^2\rangle/\omega_0.
\end{equation}

We now go to the interaction representation using canonical transformation $U_{\rm nr}(t)=T_t\exp\{-i\hbar^{-1}\int^t dt' [H_0(t')+H_{\rm b}]\}$, where $H_0(t)$ is given by eqns~(\ref{eq:H_0(t)}) and (\ref{eq:nonresonant_H_F}), $H_{\rm b}$ is the Hamiltonian of the thermal bath, and $T_t$ is the chronological ordering operator. For nonresonant modulation
\begin{equation}
\label{eq:q_t_nonresonant}
U^{\dag}_{\rm nr}(t)qU_{\rm nr}(t)\approx q_0(t)+A_{\rm osc}{}\cos\omega_Ft,\qquad A_{\rm osc}{}=A{}/(\omega_0^2-\omega_F^2).
\end{equation}
Here, $q_0(t)$ is operator $q$ in the interaction representation in the absence of the modulating field. Similarly, operator $a$ in the interaction representation is  a sum of operator $a_0(t)$ calculated for $A{}=0$ and the terms oscillating as $\exp(\pm i\omega_Ft)$; the ac-Stark shift of the eigenfrequency $\sim |\gamma| A_{\rm osc}{}^2/\omega_0$ is assumed small compared to $\omega_0$.

From eqns~(\ref{eq:quadratic_coupling}) and (\ref{eq:q_t_nonresonant}), the nonlinear coupling to the bath $H_i^{(2)}$ in the interaction representation contains a term $2q_0(t)h_{\rm b}^{(2)}(t) A_{\rm osc}{}\cos\omega_Ft$, which has the same structure as the linear coupling, eqn~(\ref{eq:linear_relaxation}), except for the time-dependent factor $\propto A_{\rm osc}{}$. Therefore the contribution of the field-induced scattering to the decay rate $\Gamma$\index{decay rate} is described by an expression similar to equation (\ref{eq:linear_relaxation}) except that one should replace $\omega_0\to \omega_0\pm \omega_F$. This gives the overall rate as $\Gamma_F=\Gamma+\Gamma_+ + \Gamma_- - \Gamma_{\rm i}$, with
\begin{eqnarray}
\label{eq:Gamma_heating}
\Gamma_{\pm}=\frac{A^2_{\rm osc}{}}{2\hbar\omega_0}{\rm Re}~\!\int\nolimits_0^{\infty}dt\langle [h_{\rm b}^{(2)}(t),h_{\rm b}^{(2)}(0)]\rangle_{\rm b}e^{i(\omega_0\pm \omega_F)t} \quad (\omega_0\pm\omega_F >0);
\end{eqnarray}
rate $-\Gamma_{\rm i}$ is formally given by eqn~(\ref{eq:Gamma_heating}) for $\Gamma_-$ in which $\omega_0-\omega_F < 0$; therefore $\Gamma_{\rm i}>0$ and $\Gamma_-=0$ for $\omega_F-\omega_0>0$, whereas $\Gamma_- > 0$ and $\Gamma_{\rm i}=0$ for $\omega_0-\omega_F>0$. The terms $\Gamma_{\pm}$ lead to the increase of the overall decay rate, whereas $\Gamma_{\rm i}$ reduces the rate and, if dominates, can make $\Gamma_F$ negative.

The mapping on the effectively linear in $a,a^{\dag}$ coupling suggests that, in the presence of nonresonant modulation, the master equation in the interaction representation\index{master equation} has the conventional form for an oscillator with no modulation, cf. eqn~(\ref{eq:master_eq}),
\begin{eqnarray}
\label{eq:master_heating}
&&\partial_t\rho = -\Gamma_F(\bar n_F + 1)(\ad_0 a_0\rho-2a_0\rho\ad_0+\rho \ad_0 a_0)-\Gamma_F\bar n_F(a_0\ad_0\rho-2\ad_0\rho a_0+\rho a_0\ad_0),\nonumber\\
&& \bar n_F=\left\{\Gamma\bar n(\omega_0)+\Gamma_+\bar n(\omega_0+\omega_F)+\Gamma_-\bar n(\omega_0-\omega_F)+\Gamma_{\rm i}\left[\bar n(\omega_F-\omega_0)+1\right]\right\}/\Gamma_F
\end{eqnarray}
where $a_0\equiv a_0(t), \ad_0\equiv \ad_0(t)$. Each relaxation mechanism in Fig.~\ref{fig:heating_cooling} gives an additive contribution to eqn~(\ref{eq:master_heating}), with the Planck number at the energy transferred to the bath in the corresponding elementary process.

The stationary solution of eqn~(\ref{eq:master_heating}) has the form of the Boltzmann distribution with temperature
\[T^* = \hbar\omega_0/k_B\ln[(\bar n_F +1)/\bar n_F].\]
This expression goes into the discussed above limits if one of the scattering mechanisms dominates, which requires that parameters $\Gamma_{\pm},\Gamma_{\rm i}$ significantly differ from each other.

The difference between parameters $\Gamma_{\pm},\Gamma_{\rm i}$ is determined by the difference between the density of states of the thermal bath at the appropriate frequencies. It can be large if the oscillator is coupled to the thermal bath via an underdamped vibrational mode, with $h_{\rm b}^{(2)}$ being proportional to the coordinate of this mode $q_{m}$ \shortcite{Dykman1978}. The mode decay rate $\Gamma_m$ should largely exceed $\Gamma,|\Gamma_F|$, but still it can be small compared to the mode frequency $\omega_m$ and $|\omega_m - \omega_0|$. One can then selectively tune $\omega_F$ to resonance and for example, for  $\omega_F\approx \omega_m -\omega_0\gg\omega_0$ achieve significant oscillator cooling.

Similar effects occur and similar description applies if the coupling to the bath is linear in $a,a^{\dag}$, eqn~(\ref{eq:linear_relaxation}), but nonresonant modulation is parametric, with Hamiltonian $H_{\rm par}$, eqn(\ref{eq:modulation_hamiltonian}), and $ \omega_F,|\omega_F-2\omega_0|\gg \Gamma,|\gamma|\langle q^2\rangle/\omega_0$. Parameters $\Gamma_{\pm}, \Gamma_{\rm i}$ in this case are also quadratic in the modulation amplitude $F$ and are determined by the correlators of $h_{\rm b},h_{\rm b}^{\dag}$ at frequencies $\omega_0\pm\omega_F$.

The possibility of reducing the vibration temperature by modulation has been attracting much attention recently, in particular in the context of optomechanics,\index{optomechanics} \shortcite{Braginsky2002,Kippenberg2008}. In an optomechanical system, the oscillator (a vibrating mirror) is coupled to a cavity mode, which is driven by external radiation. A quantum theory of cooling of the mirror in this case was developed by \shortciteN{Wilson-Rae2007} and \shortciteN{Marquardt2007}. If the incident radiation is classical, in the appropriately scaled variables the coupling and modulation are described by Hamiltonians $H_i^{(m)}$ and $H_F^{(m)}$, respectively, with
\begin{equation}
\label{eq:cavity_mirror}
H_i^{(m)}=c_mqq_m^2, \qquad H_F^{(m)}=-q_mA{}\cos\omega_Ft,
\end{equation}
where $q$ and $q_m$ are the coordinates of the mirror and the mode. In cavity optomechanics one usually writes $H_i^{(m)}=c_mqa_m^{\dag}a_m$; the following discussion immediately extends to this form of the interaction.

In the absence of coupling to the mirror the cavity mode is a linear system, hence $q_m(t) = q_{m0}(t) + [\chi_m(\omega_F)\exp(-i\omega_Ft) + {\rm c.c.}]A{}/2$, where $q_{m0}(t)$ is the mode coordinate in the absence of modulation and  $\chi_m(\omega)$ is the susceptibility of the mode \shortcite{Marquardt2007}. The coupling $H_i^{(m)}$ in the interaction representation then has a cross-term $\propto q_0(t)q_{m0}(t)\exp(\pm i\omega_Ft)$. Since the cavity mode serves as a thermal bath for the mirror, this term is fully analogous to the similar cross-term in $H_i^{(2)}$ that comes from modulation of the oscillator, with $c_mq_{m0}(t)$ playing the role of $h_b^{(2)}(t)$. One can then describe the dynamics of the mirror by eqns~(\ref{eq:Gamma_heating}) and (\ref{eq:master_heating}) in which $h_{\rm b}^{(2)}(t)$ is replaced with $c_mq_{m0}(t)$ and $A^2_{\rm osc}{}$ is replaced with $|A{}\chi_m(\omega_F)|^2$.

Another potentially important contribution to nonresonant cooling, heating, or excitation of the oscillator can come from the direct nonlinear interaction of the oscillator and the thermal bath with the modulation. For an electromagnetic modulation such interaction is due to nonlinear polarizability. For modulation  $A\cos\omega_Ft$ the corresponding interaction Hamiltonian is
\begin{equation}
\label{eq:direct_bath_osc_modulation}
H_i^{(F)}=-qh_{\rm b}^{(F)}A\cos\omega_Ft.
\end{equation}
For example, here $h_{\rm b}^{(F)}$ can be the coordinate of a comparatively quickly decaying mode of a nanomechanical resonator.

The effect of interaction (\ref{eq:direct_bath_osc_modulation}) is again described by eqns~(\ref{eq:Gamma_heating}) and (\ref{eq:master_heating}) in which $h_{\rm b}^{(2)}(t)$ is replaced with $h_{\rm b}^{(F)}(t)$ and $A^2_{\rm osc}$ is replaced with $A{}^2/4$. If more than one of the above mechanisms is relevant, in calculating the rates of modulation-induced decay one should take into account the interference terms, i.e., write the overall effective coupling in the interaction representation as $a_0(t)\exp(\pm i\omega_Ft)\tilde h_{\rm b}(t) + {\rm H.c.}$ and then express the rates $\Gamma_{\pm}, \Gamma_{\rm i}$ in terms of the commutator of  $\tilde h_{\rm b}(t)$ similar to eqn~(\ref{eq:Gamma_heating}).

Nonresonant modulation modifies the power spectrum of the oscillator \shortcite{Dykman1978,DK_review84}. Cooling of a nonlinear oscillator can lead to narrowing of its spectrum at frequency $\omega_0$ even though $\Gamma_F>\Gamma$. Unless there are symmetry constraints, the modulation can also lead to spontaneous emission of photons or phonons at frequencies $|\omega_0\pm\omega_F|$. It can also lead to amplification of a weak external field at frequency $\omega_F-\omega_0$ for $T^*>0$.

\section{Conclusions}
\label{sec:conclusions}

We described the dynamics of a modulated nonlinear oscillator for resonant and nonresonant modulation. Two types of resonant modulation were considered, an additive force with frequency $\omega_F$ close to the oscillator eigenfrequency $\omega_0$ and parametric modulation with frequency $\omega_F$ close to $2\omega_0$. For resonant modulation, quantum dynamics in the rotating frame is characterized by three parameters, the scaled modulation intensity, decay rate, and Planck constant, which are given in Table~\ref{table:parameters}. Of primary interest was the parameter range where the oscillator displays bistability of forced vibrations. Fluctuations were assumed small on average, so that the smearing of the classically stable states in phase space is small compared to the interstate distance. For $T>0$ the system lacks detailed balance

Relaxation of the modulated quantum oscillator is accompanied by quantum noise. It leads to a finite-width distribution over quasienergy states even for $T\to 0$. We found the distribution for weak damping, where the width of the quasienergy levels is small compared to the level spacing. Near its maximum the distribution is Boltzmann-like. The far tail is of non-Boltzmann form. It determines the exponent in the rate of switching between the stable vibrational states. The switching occurs via transitions over the effective barrier that separates the states in phase space. We called this quantum activation and studied the corresponding effective activation energy. Remarkably, even for $T\to 0$ it is smaller than the exponent for switching via tunneling. Therefore interstate switching occurs via quantum activation, not tunneling, unless the decay rate is exponentially small. We also found, using a different method, the switching rate close to bifurcation points. Its scaling with the parameters is given in Table~\ref{table:bifurcation}.

We found the power spectra and the susceptibility of resonantly modulated oscillators. The spectra have a characteristic shape, which depends on the interrelation between the decay rate, the quasienergy level spacing, and the nonequidistance of the quasienergy levels. Where the nonequidistance exceeds the level width, the spectra display a fine structure, which sensitively depends on the effective temperature of the quasienergy distribution near its maximum. They can also display a characteristic supernarrow peak where the stationary populations of the coexisting vibrational states are close to each other.

An interesting effect of nonresonant modulation of the oscillator is that it can significantly change the oscillator distribution over the Fock states, leading to heating, cooling, or excitation of self-sustained vibrations depending on the modulation frequency and the coupling to the thermal bath. We show that different coupling and modulation mechanisms can be described in a similar way and that the distribution can remain of the Boltzmann form in a broad range of oscillator energies.

This research was supported in part by DARPA through the DEFYS program and by the NSF, Grant EMT/QIS 089854.




\printindex
\end{document}